\documentclass[aps,pra,showpacs,reprint,10pt]{revtex4-2}

\usepackage{physics,amsmath,amssymb}
\usepackage{dcolumn}
\usepackage{bm}
\usepackage{graphicx}
\usepackage{hyperref}
\usepackage{tikz}
\usetikzlibrary{shapes}
\usepackage{etoolbox}

\definecolor{donkerblauw}{RGB}{0, 114, 189}
\definecolor{mediumdonkerblauw}{RGB}{19, 159, 255}
\definecolor{mediumlichtblauw}{rgb}{0.3, 0.75, 0.93}
\definecolor{lichtblauw}{RGB}{108, 176, 244}
\definecolor{grijs}{rgb}{0.5, 0.5, 0.5}
\definecolor{fluoblauw}{RGB}{15,255,255}

\newcommand{\noiseone}{\raisebox{0.5pt}{\tikz{\node[draw=black,scale=0.38,regular polygon, regular polygon sides=4,fill=black](){};}}}
\newcommand{\noisethree}{\raisebox{0.5pt}{\tikz{\node[draw=donkerblauw,scale=0.38,diamond,fill=donkerblauw](){};}}}
\newcommand{\benchmark}{\raisebox{0.5pt}{\tikz{\node[draw=grijs,scale=0.4,circle,fill=grijs](){};}}}
\newcommand{\noisefour}{\raisebox{0.5pt}{\tikz{\node[draw=mediumdonkerblauw,scale=0.38,regular polygon, regular polygon sides=3, fill=mediumdonkerblauw,rotate=0](){};}}}
\newcommand{\noisefive}{\raisebox{0.5pt}{\tikz{\node[draw=mediumlichtblauw,scale=0.35,regular polygon, regular polygon sides=3,fill=mediumlichtblauw,rotate=180](){};}}}

\begin{document}

\preprint{APS/123-QED}

\title{Non-equilibrium steady states and critical slowing down in the dissipative Bose-Hubbard model}
\author{Robbe Ceulemans}
\email{robbe.ceulemans@uantwerpen.be}
\affiliation{TQC, Universiteit Antwerpen, Universiteitsplein 1, 2610 Antwerpen, Beglium}
\author{Michiel Wouters}
\affiliation{TQC, Universiteit Antwerpen, Universiteitsplein 1, 2610 Antwerpen, Beglium}
\date{\today}

\begin{abstract}
Motivated by recent experiments, we study the properties of large Bose-Hubbard chains with single-particle losses at one site using classical field methods.
We construct and validate a compact effective model that reduces computations to only a few sites. We show that in the mean-field approach the description captures the stationary states of the dissipative mode very well. Not only is there a good quantitative agreement in the hysteresis loop, the dark soliton state can be reproduced as well due to the preservation of the $U(1)$ symmetry. Bimodality of the steady states, observed on longer timescales, is studied using the truncated Wigner method. We compare the switching statistics and derive the effective Liouvillian gap in function of the tunneling, showing that the effective description underestimates fluctuations.
\end{abstract}
\maketitle

\section{\label{sec:Intro} Introduction}
In practice quantum systems are subject to dissipation, leading to decoherence and loss of entanglement. Studies in recent years have however shown that for some systems a well chosen coupling to its environment can have a beneficial effect and drive it to desired highly entangled states \cite{diehl_quantum_2008,verstraete_quantum_2009,barreiro_open-system_2011}. Combining such engineered dissipation with an internal or external driving has led in a variety of systems to exotic non-equilibrium steady states (NESS) that can not be reached in closed systems \cite{drummond_quantum_1980,wouters_excitations_2007,kessler_dissipative_2012,kordas_dissipation-induced_2012,casteels_power_2016,yanay_reservoir_2018,huybrechts_dynamical_2020,wang_pattern_2020}. Properties of these NESS often greatly differ from the thermal equilibrium states of the Hamiltonian.


While dissipation times in photonic systems are often similar to other relevant timescales \cite{walls_quantum_2008,carusotto_quantum_2013,noh_quantum_2017,verstraelen_gaussian_2018,pieczarka_observation_2020}, the intrinsic losses in configurations of ultracold atoms are slow on their characteristic timescales. They offer in general all-round controllable set-ups, well isolated from the environment and with a great adaptability of the microscopic parameters through external fields. Additional losses can therefore be introduced by externally engineering dissipation, giving good control over the relative importance of dissipative and Hamiltonian dynamics. One particular experimental implementation of a lossy atomic system was realized on a cigar-shaped BEC, tightly confined along $x$- and $y$-axis, in a periodic potential along the $z$-direction \cite{labouvie_negative_2015,labouvie_bistability_2016,benary_experimental_2022}. Particle losses around one potential minimum in the centre were induced by ionizing atoms with a focused electron beam. Tunneling from a large sequence of highly occupied wells towards the dissipative site provide an effective drive that leads to long-lived steady states. Evidence for a first order phase transition was observed in this setup.

\begin{figure}[h]
    \centering
    \includegraphics[width=8.6cm]{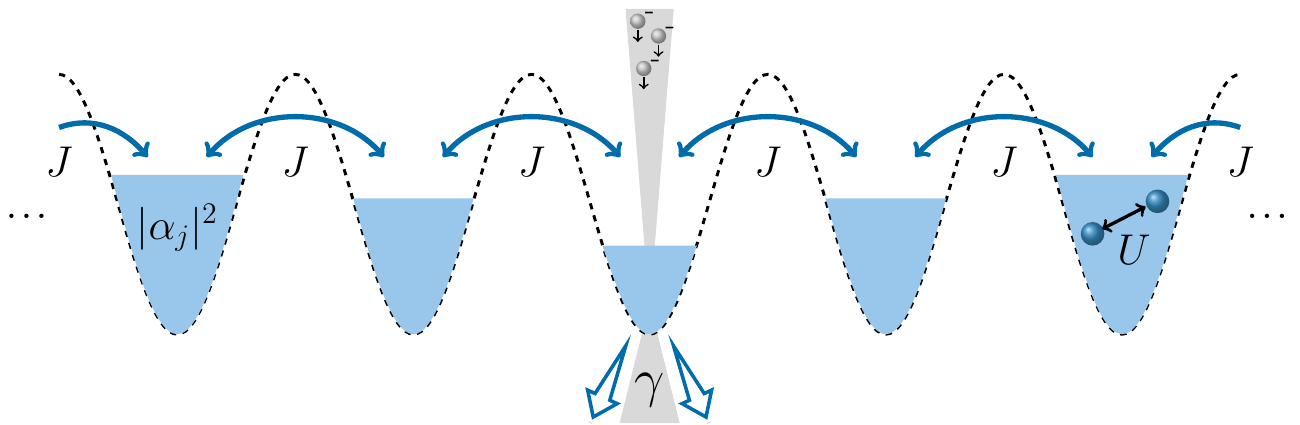}
    \caption{Schematics of the experiment. Losses at a tunable rate $\gamma$ take place on a single site of a BEC loaded into a 1D optical lattice. For sufficiently large lattice depths, only nearest-neighbour tunneling and on-site interactions contribute significantly to the dynamics. The highly occupied modes left and right of the dissipation act as reservoirs inducing large particle streams to account for the losses.}\label{Fig:Schematic}
\end{figure}

Theoretically, Reeves and Davis have pioneered the modeling of this experimental setup within a classical field description. Because the full 3D system is numerically quite involved, they have reduced the modeling to a few lattice sites along the $z$-direction while keeping the full two-dimensional structure in the transverse direction \cite{reeves_bistability_2023}, demonstrating for example a close analogy with a resonantly driven nonlinear optical cavity \cite{drummond_quantum_1980}. A detailed understanding of the validity of the reduction in the $z$-direction is however missing. In this work, we wish to fill this gap by considering the complementary problem of a 1D Bose-Hubbard chain with losses in the central site, shown schematically in Fig.~\ref{Fig:Schematic}. This system is simple enough to allow for a full numerical study within the truncated Wigner approximation (TWA) and therefore constitutes a good starting point for the development of models that truncate the number of lattice sites. The aim of this work is the construction of a minimal effective description that captures not only the same steady-state physics, but also the dynamics and quantum fluctuation induced switches between the branches of the bistability \cite{benary_experimental_2022}.

The problem of localized particle loss in small Bose-Hubbard chains has been studied extensively over the past years. Spontaneous symmetry breaking is reported in photonic dimers with additional coherent driving \cite{garbin_spontaneous_2022,giraldo_driven-dissipative_2020,giraldo_semiclassical_2022}. Discrete breathers have been mainly analysed in trimers and small extended arrays \cite{kordas_dissipation-induced_2012,witthaut_beyond_2011,kordas_non-equilibrium_2015}. Larger chains with dissipation on one site have been studied for small occupations in the Mott-insulator regime \cite{barmettler_controllable_2011,kiefer-emmanouilidis_current_2017}, for large particle numbers in the non-interacting limit \cite{kepesidis_bose-hubbard_2012} and the weakly interacting regime \cite{kordas_decay_2013,bychek_probing_2019}, and recently even for spinless fermions \cite{van_nieuwenburg_dynamics_2018,froml_ultracold_2020}. The continuous counterpart has also been topic of theoretical studies \cite{brazhnyi_dissipation-induced_2009,zezyulin_macroscopic_2012,sels_thermal_2020}. 

The structure of this paper is as follows. In Sec.~\ref{sec:models} we introduce the dissipative Bose-Hubbard model with single-particle losses at one site. In Sec.~\ref{sec:MF} we study this system in a mean-field framework, discussing the apparent bistability. Based on these results an effective description is introduced that shows great quantitative agreement for the dissipative site. In Sec.~\ref{sec:Solitons} we study the formation of a dark soliton fixed in position by the dissipation appearing in both these models. In Sec.~\ref{sec:QFl} quantum fluctuations are taken into account using the TWA, which captures the sudden switches between steady states. We perform a study of the characteristic timescales of this effect and the closing of the Liouvillian gap that is inseparably linked to this. Finally, in Sec.~\ref{sec:Conclusion} we summarize our results.

\section{\label{sec:models} The Bose-Hubbard model with local dissipation}
A BEC loaded into a 1D periodic potential can, for large lattice depths and tight trapping in the transverse directions, be well approximated by the Bose-Hubbard model given by \cite{bloch_many-body_2008}
\begin{equation}
    \hat{H}_{BH} = -J\sum_{j=1}^{L-1}\big(\hat{a}_{j+1}^\dagger\hat{a}_j + h.c.\big) + \frac{U}{2}\sum_{j=1}^L\hat{a}^\dagger_j\hat{a}^\dagger_j\hat{a}_j\hat{a}_j.
\end{equation}
Here $\hat{a}_j^\dagger$ and $\hat{a}_j$ denote the bosonic creation and annihilation operators at the j-th site, $J$ the nearest-neighbour tunneling amplitude and $U$ the on-site interaction energy. 

When a quantum system is coupled to a Markovian environment, the dynamics of its density matrix is governed by the Lindblad master equation~\cite{breuer_theory_2002}:
\begin{equation}
    \pdv{\hat{\rho}}{t} = \mathcal{L}\hat{\rho} = -\frac{i}{\hbar}\left[\hat{H},\hat{\rho}\right] - \sum_j \frac{\gamma_j}{2}\left(\left\{\hat{\Gamma}_j^\dagger\hat{\Gamma}_j,\hat{\rho}\right\} - 2\hat{\Gamma}_j^\dagger\hat{\rho}\hat{\Gamma}_j\right).\label{Eq:LindbladMaster}
\end{equation}
Here the $\hat{\Gamma}_j$ are the quantum jump operators that represent the effect of the coupling to the environment. For the case of localized atomic losses on the central cite ($j=c$), there is a single jump operator given by $\hat \Gamma = \hat  a_c$ and $\gamma_j=\gamma\delta_{jc}$.

This system only features losses and no compensating driving. The true steady state of the system at infinite time is therefore trivially empty. At intermediate times however, the modes surrounding the central site will act as a reservoir making the lossy site effectively a driven-dissipative system. Competition between the losses and the Bose-Hubbard dynamics, that tends to level the particle number in all sites, drives the system in a good approximation to a NESS \cite{labouvie_bistability_2016,reeves_bistability_2023,benary_experimental_2022}. In the following, we will call the quasi steady state at intermediate times simply the steady state of the system, the time scale over which this state exists becoming longer when increasing the system size $L$ and tending to infinity in the thermodynamic limit.

\section{\label{sec:MF} Mean field description}
Typically, quantum fluctuations play an important role in 1D systems \cite{witthaut_beyond_2011,barmettler_controllable_2011,kordas_non-equilibrium_2015,kiefer-emmanouilidis_current_2017}. An appropriate description of this system taking into account fluctuations is therefore given in Sec.~\ref{sec:QFl}. For weak interactions and large particle numbers however, the main features of the steady state can be understood within a mean field description, where each site is assumed to be in a coherent state. Within this approximation, the master equation reduces to a discrete Gross-Pitaevskii equation (GPE) for the coherent field amplitudes $\alpha_j = \langle\hat{a}_j\rangle$:
\begin{equation}
\begin{split}
    i\hbar\dv{}{t}\alpha_j = -J&\big(\alpha_{j-1} + \alpha_{j+1}\big) + U\big(|\alpha_j|^2-1\big)\alpha_j\\ &- i\frac{\gamma\delta_{c,j}}{2}\alpha_j.
    \end{split}\label{Eq:GP_BoseHubbard}
\end{equation}
The complex amplitudes are numerically time evolved according to this GPE using the \textit{DifferentialEquations.jl} package from the Julia programming language \cite{rackauckas_differentialequationsjl_2017}. Calculations are performed for large chains ($L\sim 10^{2}$) deep in the superfluid regime for an initial occupation per site $n_0=700$.

\subsection{\label{subsec:Bistability} Bistability}
\begin{figure}[h]
    \centering
    \includegraphics[width=8.6cm]{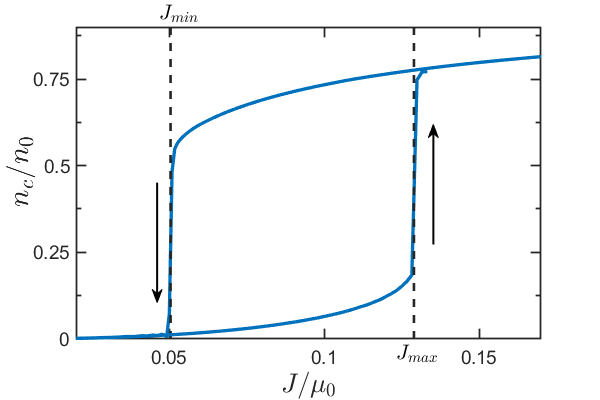}
    \caption{Normalized steady-state occupation of the central dissipative mode in the Bose-Hubbard array at different tunneling strengths $J$ with a fixed dissipation rate $\gamma/\mu_0=0.21$. Starting at $J<J_{\text{min}}$ ($J>J_{\text{max}}$) and increasing (decreasing) the tunneling strength the solution follows the lower (upper) branch. A sudden discontinuous jump to the opposing branch occurs when the bistable regime is left.}\label{Fig:Hysteresis}
\end{figure}

In analogy with experimental observations \cite{labouvie_bistability_2016}, the mean field theory predicts bistable behavior, illustrated in Fig.~\ref{Fig:Hysteresis}, where we show the relative particle number of the lossy site for an adiabatic ramping up and down of the tunneling rate. 
At zero tunneling rate, the central site is decoupled, such that the dissipation will simply empty it within a time of the order of $\gamma^{-1}$. When on the other hand the tunneling is very large, particle currents can easily compensate for the losses, resulting in a steady state with large occupation.
At intermediate tunneling rates, the occupation depends on the system history. When starting from a central site with low occupation, the large interaction energy difference between the central site and its first neighbour prevents tunneling in the same way as in the self-trapping regime of the bosonic Josephson junction~\cite{smerzi_quantum_1997,meier_josephson_2001,albiez_direct_2005,secli_signatures_2021}. Modes on both sides of the dissipative system remain mostly undepleted.
When, on the other hand, starting from a central site with the same occupation as its neighbours, the tunneling is efficient and the relative occupation remains close to one. 

With a slow adiabatic increase of $J$ starting from zero, the system will move along the lower stable branch. The opposite happens when starting at large tunneling strengths and sweeping down along the upper stable branch. For large and small values of $J$ both branches overlap, but for $J_{min}<J<J_{max}$ the system is bistable.

The hysteretic behavior is in direct analogy with that of the coherently driven nonlinear resonator, that is at the mean field level described by \cite{drummond_quantum_1980}
\begin{equation}
    i\hbar\dv{}{t} \alpha_c = F e^{-i\omega_d t}+ U |\alpha_c|^2 \alpha_c-i\frac \gamma 2 \alpha_c,
    \label{Eq:kerr}
\end{equation}
where $\omega_d$ is the driving frequency. The connection between the Kerr model~\eqref{Eq:kerr}, that can be solved analytically for the NESS, and our Bose-Hubbard system is readily made by neglecting the backaction of the central site on its nearest neighbours. The amplitude of the neighbouring sites is then equal to $\sqrt{n_0}$ and their frequency is given by the chemical potential $\mu_0=U n_0$ \cite{pitaevskii_bose-einstein_2016}, such that $\alpha_{c\pm 1} = \sqrt{n_0} e^{-i \mu_0 t}$. Substituting in Eq.~\eqref{Eq:GP_BoseHubbard} yields exactly Eq.~\eqref{Eq:kerr} with $F=2J\sqrt{n_0}$ and $\omega_d =\mu_0$.

\subsection{\label{subsec:IncohPump} Incoherent pumping model}
The simplification of the dynamics from the full GPE to the coherently driven nonlinear resonator is particularly attractive in view of modeling the experiments in Refs.~\cite{labouvie_bistability_2016,benary_experimental_2022} that consist of an array of two-dimensional gases, and worked out by Reeves and Davis \cite{reeves_bistability_2023}, since it then allows to reduce the dynamics of a three-dimensional to a two-dimensional system.

\begin{figure}[h]
    \centering
    \includegraphics[width=8.6cm]{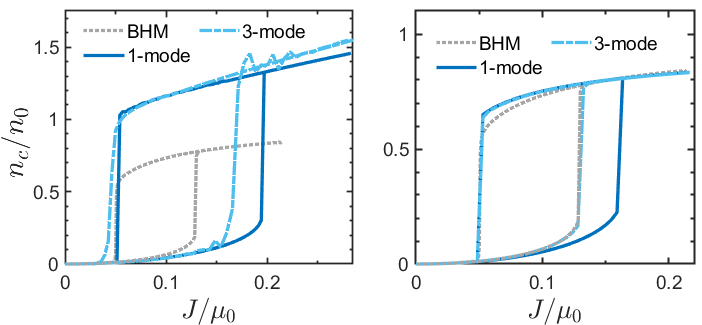}
    \caption{Comparison of the lossy site steady-state occupations in the BHM with those from the effective descriptions in Eq.~\eqref{Eq:kerr} (a) and Eq.\eqref{Eq:IncohDrive} (b). (a) The coherently driven single mode, although in qualitative agreement, grossly overestimates the upper bound on the bistability regime and predicts overfilling in the upper branch. With increasing size (multiple modes with coherent pumping at the edges) the boundaries of the bistability region shift in favour of the Bose-Hubbard simulation benchmark, but overfilling becomes even more pronounced. (b) A three-well system with incoherent drive at the edges is already a much better approximation, with a remarkable overlap of the stable upper branch. A slightly larger five-well system also brings the bistability bounds in quantitative agreement.}\label{Fig:Compare}
\end{figure}

While at the qualitative level, there is a good correspondence between the GPE model of the whole chain and the coherently driven resonator, there are significant quantitative differences in the shape of the hysteresis as is visible in Fig.~\ref{Fig:Compare}(a). The most salient discrepancies are the overestimation of the upper bistability threshold and the density on the upper bistability branch. The latter is in particular unphysical, because the coherently driven model predicts a \textit{larger} occupation on the dissipative site than on the neighbouring sites. 

In order to obtain a more accurate reduced description for a Bose-Hubbard system with local dissipation, we develop a model that is inspired by descriptions of exciton-polariton condensates, where the losses are compensated by the gain from an incoherent reservoir \cite{wouters_excitations_2007,carusotto_quantum_2013}. 
From the above discussion on the bistability, it is clear that the resonant tunneling between the lossy site and its neighbours is an essential ingredient of the dynamics. Keeping the amplitude of the first neighbours fixed leads to artefacts such as overfilling. We therefore want to treat them as dynamical variables, while approximating the contribution from further away modes by a single driving term. In order to maintain the $U(1)$ symmetry, we model them as incoherent pumping baths.

\begin{figure}[h]
    \centering
    \includegraphics[width=8.6cm]{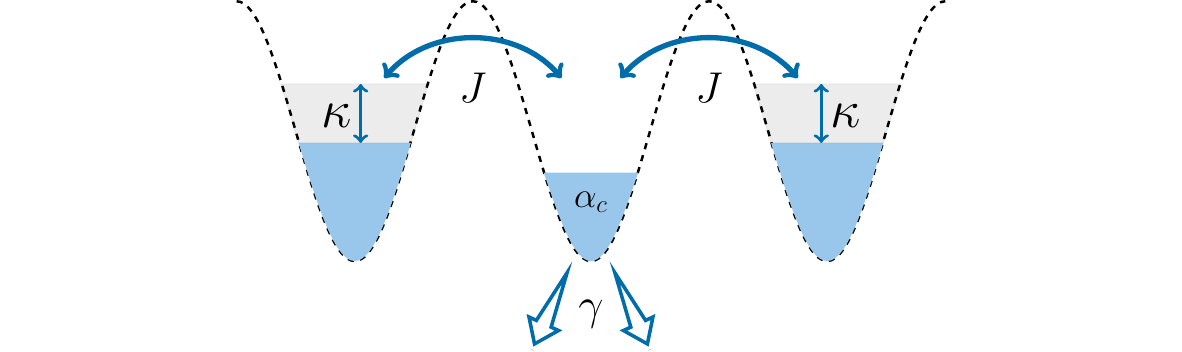}
    \caption{A schematic representation of the compact incoherently driven model. This Bose-Hubbard trimer with losses in the centre has boundary driving in the form of a saturation with maximum occupation $n_0$, which mimics the refilling from a large sequence of modes with the same occupation.}\label{Fig:IncohDrive}
\end{figure}

We start by looking at a small trimer ($L=3$), schematically shown in Fig.~\ref{Fig:IncohDrive}, with losses in the central mode, $c=2$, and boundary driving
\begin{equation}
\begin{split}
    i\hbar\dv{}{t}\alpha_{1,3} = -J&\left(\alpha_{2} - \alpha_{1,3}\right) + U|\alpha_{1,3}|^2\alpha_{1,3}\\ &+ i\frac{\kappa}{2}\left[1-\frac{|\alpha_{1,3}|^2}{n_0}\right]\alpha_{1,3}.
    \end{split}\label{Eq:IncohDrive}
\end{equation}
Here $\kappa$ is the rate of the saturation, simulating refilling from a large number of highly occupied wells with mean occupation $n_0$. A gain term like this yields a typical refilling that follows a logistic curve. This is shown to be the best fit for the refilling in a Bose-Hubbard array \cite{labouvie_negative_2015,fischer_models_2017,mink_variational_2022}. The on-site tunneling that is added in Eq.~\eqref{Eq:GP_BoseHubbard} simply rescales the groundstate energy.

Steady states of the dissipative mode in this configuration are shown in Fig.~\ref{Fig:Compare}(b), in comparison with the solutions from the full chain. It is clear that the three-mode incoherent pumping model outperforms the resonant pumping one. Most significant is the similarity of the upper branch for which the behaviour depends mainly on the value of $\kappa$. Tuning the refilling rate we find that the best agreement is obtained for $\kappa=c_s=\sqrt{2JUn_0}$, exactly the speed of sound in a Bose-Hubbard chain. The lower branch is less affected by variations in the refilling rate.

Even better agreement with the benchmark is obtained by slightly increasing the size. We take now $L=5$ coupled modes where the first and last are again saturated as described in Eq.~\eqref{Eq:IncohDrive}. The results are shown in Fig.~\ref{Fig:Compare}(b) with the dashed line, showing overall very good agreement with the complete system.

\section{\label{sec:Solitons} Discrete soliton formation}
Beside the two symmetric solutions with large and small occupation of the lossy site described above, a third asymmetric state of the system can be reached. It displays a $\pi$ phase difference between the left and right neighbours of the lossy site such that the density in the lossy site is exactly zero, owing to the destructive interference between the two tunneling currents. Because the density in the lossy site then vanishes exactly within the mean field description, it is not affected by the dissipative part of the dynamics and it is, at the mean field level, a dark state of the Liouvillian.

\begin{figure}[h]
    \centering
    \includegraphics[width=8.6cm]{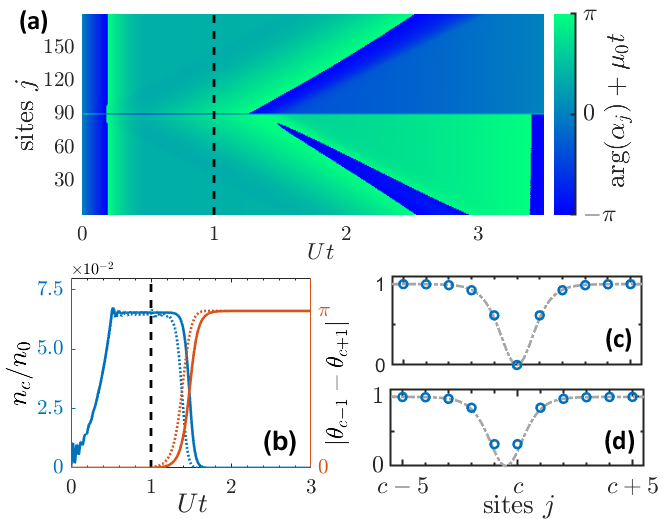}
    \caption{Formation of a standing dark soliton. (a) The argument of the coherent field amplitudes $\alpha_je^{i\mu_0 t/\hbar}$ in function of time show a phase separation that breaks the mirror symmetry after small random fluctuations are added at $U t_p=1$ (dashed line). Reservoir modes on both sides go from rotating in phase to antiphase. (b) As a consequence a phase difference of $\pi$ builds up between the left and right neighbours of the lossy site. Driving from both sides interferes destructively, causing the occupation in the central mode to vanish. The effective model (dotted lines) captures this behaviour very well with only a small difference in the time it takes for this transition to occur. (c), (d) Normalized field amplitude at a few sites around the centre either with losses at $j=c$ (c) or without losses, but with an initial $\pi$ phase jump imprinted on the chain (d). Both configurations are appearances of the discrete dark soliton from Eq.~\eqref{Eq:DiscDarkSol} with a shift of $m$ by half a lattice constant (dashed lines).}\label{Fig:Soliton}
\end{figure}

Fig.~\ref{Fig:Soliton} shows the dynamics of the formation of the dark state starting from a condensate with uniform phase, zero tunneling and an entirely empty lossy site. The tunneling strength is then slowly increased to reach the NESS on the lower bistability branch for $J/\mu_0=0.1$. At time $t_p$ a small phase perturbation is added manually to the mean-field dynamics in the form of a multiplicative phase $e^{i\pi\times 0.01 \xi_j}$, where $\xi_j$ are normally distributed with unit variance. Following this perturbation, the modes on the left and right move out of phase to the point of perfect destructive interference in the centre where the particle number drops to zero. A discrete dark soliton is formed with the zero amplitude minimum locked at the dissipative mode.

The above numerical analysis suggests that the symmetric state on the lower branch of the bistability is not stable with respect to small phase perturbations. This is confirmed by a linear stability analysis of our incoherent pumping model. As shown in Fig. \ref{Fig:Soliton} (b) with the dotted line, the effective description (for $L=5$) reproduces the transition from the lower bistability branch to the dark state, with only a small discrepancy in the transition time. It is worth pointing out that it is the $U(1)$ invariance of the incoherent pumping model that allows to describe the transition to the dark soliton. This is in contrast to the coherent pumping model where the phase of the frequency of the drive is fixed externally and no spontaneous phase dynamics takes place.

The stationary dark state is, far from the edges, well described by \cite{kivshar_dark_1994,brazhnyi_dissipation-induced_2009}
\begin{equation}
    \alpha_j = \sqrt{n_0}\tanh\big[\sqrt{\mu_0/2J}(x_j-m)\big]\exp{-i\mu_0t/\hbar},\label{Eq:DiscDarkSol}
\end{equation}
where $m=c$ is the soliton location, as can be seen in the spatial amplitude profile in Fig.~\ref{Fig:Soliton}(c). As a comparison we show the stationary state, after imaginary time evolution, for $\gamma=0$, but with an initial $\pi$ phase jump at $j=c$ in  Fig.~\ref{Fig:Soliton} (d). This state coincides with Eq.~\eqref{Eq:DiscDarkSol} when $m=c-1/2$. These two configurations are referred to as an on-site and inter-site dark soliton respectively. They can be viewed as realizations of the same soliton state translated through the lattice by half a lattice constant \cite{kivshar_dark_1994}. Due to the energy difference brought about by the discreteness, a barrier exists between both configurations. The closed system will generally, after appropriate phase imprinting \cite{carr_dark-soliton_2001,mishmash_quantum_2009,mishmash_quantum_2009-1}, end up with a lower energy inter-site soliton, where two nearest neighbours have a $\pi$ phase difference, but no sites are completely empty. Instead, when dissipation is turned on the state with a dark site is favoured.

The instability of the lower branch towards the dark soliton state has not been observed experimentally, possibly due to the spatial extent of the condensate at each lattice site that is neglected in our 1D BHM. In what follows we assume our system to be symmetric with respect to the dissipative site, effectively disregarding the soliton state, allowing to put the focus on the steady states in the hysteresis loop.

\section{\label{sec:QFl} Quantum fluctuations}
So far, we have restricted our theoretical description to the mean field approximation. While this is sufficient to understand the classical bistable behavior, it fails to capture quantum fluctuations on top of the classical dynamics. The first correction to the classical behavior manifests itself by the switching between the two branches in the bistability region. Where these steady states are stable in the classical mean field limit, in reality they are only metastable, as recently observed experimentally \cite{benary_experimental_2022}. Tunneling between these metastable states characterizes the relaxation of the density matrix to its true unique stationary value \cite{drummond_quantum_1980,huybrechts_dynamical_2020,thorwart_dynamical_1997,thorwart_quantum_1998,rodriguez_probing_2017}.

A successful approximation that allows for the inclusion of quantum fluctuations in a tractable way and that is accurate for weakly interacting Bose gases in the quantum degenerate regime is the truncated Wigner approximation \cite{sinatra_truncated_2002}. This method is based on the Wigner distribution $W(A,A^*)$ over classical phase space parameterized by the coherent state amplitudes $A=(\alpha_1,\alpha_2,\ldots)$.
Starting from Eq.~(\ref{Eq:LindbladMaster}) an equivalent equation of motion for the Wigner function can be derived \cite{gardiner_quantum_2004}. The approximation consists of neglecting third order derivatives in this differential equation, eventually leading to a set of Langevin equations for the phase space variables, as detailed in Appendix \ref{AppSec:TWA}:
\begin{equation}
\begin{split}
    i\hbar\dv{}{t}\alpha_j = &-J(\alpha_{j-1} + \alpha_{j+1}) + U\left(|\alpha_j|^2-1\right)\alpha_j\\ &- i\frac{\gamma_j}{2}\alpha_j + \sqrt{\gamma_j/2}\xi(t).
    \end{split}\label{Eq:BoseHubbardSDE}
\end{equation}

On the one hand quantum fluctuations enter the dynamics due to the non-deterministic nature of the initial conditions $\alpha_j(t=0)$. These values are sampled from the Wigner function representing the initial state of the system (see Appendix \ref{AppSec:TWA}) and subsequently time-evolved according to Eq.~\eqref{Eq:BoseHubbardSDE}. On the other hand, associated with the dissipation is the normalized complex Gaussian noise $\xi(t)$ for which holds that
\begin{equation}
    \langle\xi(t)\rangle=0~~\text{and}~~\langle\xi(t)\xi^*(t^\prime)\rangle = \delta(t-t^\prime).
\end{equation}
Moments of the Wigner function correspond to expectation values of the Weyl-ordered products of the corresponding sets of particle operators,
\begin{equation}
    \langle\alpha_j\alpha_k\dots\alpha_l^*\alpha_m^*\dots\rangle_W = \left\langle\left\{\hat{a}_j\hat{a}_k\dots\hat{a}^\dagger_l\hat{a}^\dagger_m\dots\right\}_{sym}\right\rangle.\label{Eq:Moments}
\end{equation}
Here the subscript $W$ denotes averaging over a large ensemble of stochastic Wigner trajectories.

\subsection{Branch switching}
\begin{figure}[h]
    \centering
    \includegraphics[width=8.6cm]{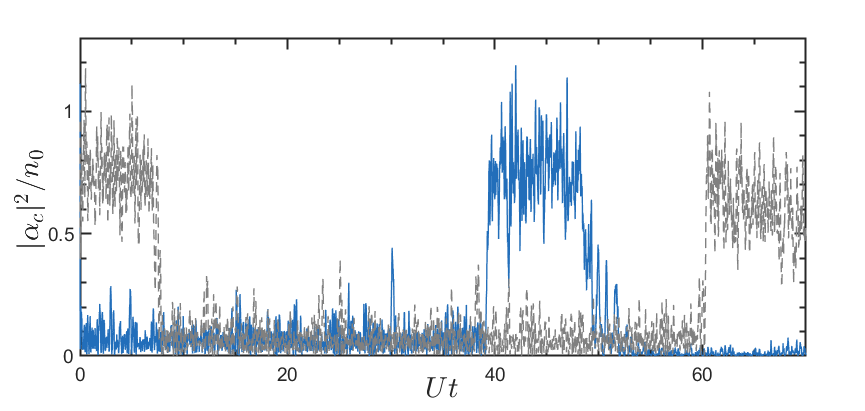}
    \caption{Normalised central mode amplitude for two single stochastic realizations of Eq.~\eqref{Eq:BoseHubbardSDE} with an initially empty (blue) or filled (grey) lossy site. The trajectories feature relatively long transient times where the amplitude fluctuates around one of the steady states, with sudden switches at seemingly random times.}
    \label{Fig:Switching}
\end{figure}

The main effect of including quantum fluctuations is shown in Fig.~\ref{Fig:Switching}, where two stochastic realizations of the central site occupation are plotted in time. Within the bistable parameter regime, this occupation number is seen to initially waver around one of the two NESS, depending on the chosen initial condition, but jumps to the complementary state can occur on longer timescales. Adding fluctuations changes the bistability to bimodality, i.e. the system probes two regions in phase space that are centered around the mean-field solutions. In principle, only averages over large ensembles correspond to quantum mechanical expectation values of observables that would allow to make comparisons with experimental observations. However, the independent Wigner trajectories already resemble single experimental measurements of the system performed for example by Benary et al. \cite{benary_experimental_2022}. 

\begin{figure}[h]
    \centering
    \includegraphics[width=8.6cm]{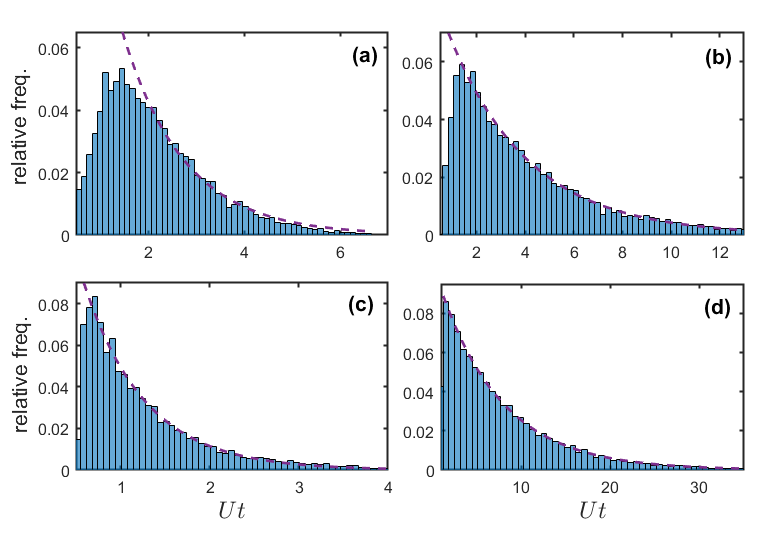}
    \caption{Histograms of the time between switches in density of the central site from high occupation to low occupation (a)-(b) and vice-versa (c)-(d).  We show examples for the dissipative BHM (a),(c) at resp. tunneling $J/\mu_0=0,07$ and $J/\mu_0=0,11$ and for the effective five-mode description (b),(d) at resp. tunneling $J/\mu_0=0,06$ and $J/\mu_0=0,12$. All distributions are characterized by a long exponential tail. At early times in the BHM switches from upper to lower branch are suppressed. This effect, although still noticeable, is less pronounced in the effective model.}\label{Fig:SwitchHist}
\end{figure}

In order to analyse the switching between the metastable branches quantitatively, we have collected statistics for the waiting time between switches from low to high atom number and vice versa. Representative examples are shown in Fig.~\ref{Fig:SwitchHist}, where simulations for the full BHM (a,c) and the effective five-site incoherent pumping model (b,d) are compared.

For the dissipative BHM, the switching time distribution deviates from an exponential decay that one would expect for a uniform Poisson process, indicating that the jumps are not completely independent. At early times suppression of the branch switching occurs. This effect is more pronounced for switches from large to small occupation, indicating the presence of dynamics preceding such a transition. The same behaviour is to a lesser extent also observable in the temporal distributions of the incoherently driven model. 

The long exponential tails determine the characteristic switching times $\tau_{up}$ and $\tau_{down}$. The time it takes the system to relax to its unique steady state, through the process of branch switching, is then characterised by  $\tau^{-1} = \tau_{up}^{-1} + \tau_{down}^{-1}$.

\subsection{\label{subsec:critical} Critical slowing down}
From the quantum Liouvillian master equation perspective, the  asymptotic decay rate of the density matrix towards the true stationary value corresponds to the inverse of the Liouvillian gap $\lambda$, defined as $\lambda = \abs{\operatorname{Re}\{\lambda_1\}}$, with $\lambda_1$ the eigenvalue from the complex spectrum with largest nonzero real part \cite{minganti_spectral_2018}. The stationary state towards which the system relaxes corresponds to the $\lambda_0=0$ eigenstate. A first order phase transition features a closing of this Liouvillian gap and thus a level touching in the eigenvalue spectrum. This results, in the vicinity of the critical point where the relaxation time diverges, at finite times in an apparent bistability.

\begin{figure}[h]
    \centering
    \includegraphics[width=8.6cm]{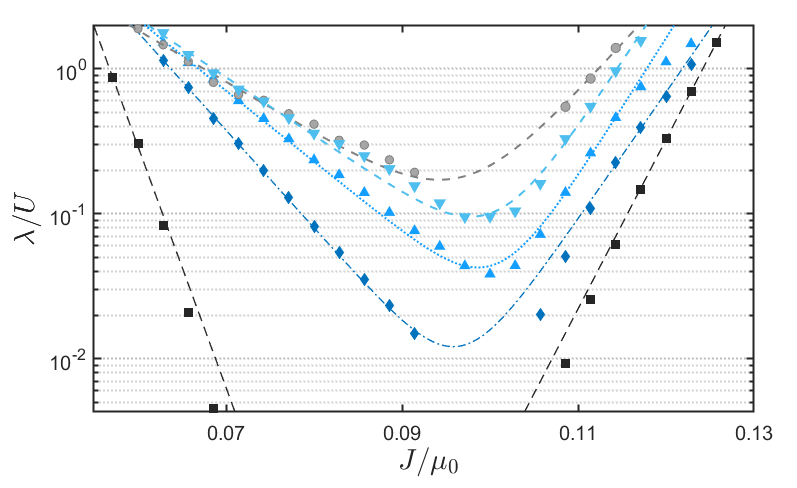}
    \caption{The effective Liouvillian gap $\lambda = \tau_{up}^{-1} + \tau_{down}^{-1}$ in function of the tunneling strength $J/\mu_0$. A large discrepancy can be noticed between the BHM (\protect\benchmark), where $\lambda$ converges to a finite value everywhere, and the incoherent pumping model (\protect\noiseone), where a true critical slowdown is observed. Increasing the noise at the edges, coming from the gain, by a factor three (\protect\noisethree), four (\protect\noisefour)and five (\protect\noisefive) we see that the minimum increases. With this modification the solution for the BHM is approached.}
    \label{Fig:LiouvillianGap}
\end{figure}

Since the Liouvillian gap determines the longest relaxation time in the dynamics we can extract its value from the analysis of the switching statistics by taking $\lambda=1/\tau$. In Fig.~\ref{Fig:LiouvillianGap} we plotted $\lambda/U$ in function of the tunneling strength in the regime where mean-field theory predicts bistability. Values from the BHM are compared to the effective model, revealing a substantial discrepancy. The minimum, that indicates the critical point $J_c$, is many orders of magnitude smaller. This difference could be due to an effective reduction of the noise in the spatially smaller system, where the dissipative site interacts with a much smaller number of modes. In a large array the many reservoir modes not only provide a saturation effect, but also brings  additional fluctuations and thus faster branch switching. The effective description does not capture this influence properly. This is confirmed when manually increasing the noise input at the edges due to the incoherent pumping. The sharpness of the dip in $\lambda(J)$ can be reduced noticeably and the BHM result more closely approached. Adding more fluctuations naturally decreases switching times, and also leads to suppression of switches at early times like in Fig.~\ref{Fig:SwitchHist}(a).

\section{\label{sec:Conclusion} Conclusion}
We have studied a 1D Bose-Hubbard chain with single-particle losses at the central site and made the comparison with a compact effective description that replaces a large number of reservoir modes by a single saturation term. In the mean-field approach, we found good quantitative agreement of the steady-state population of the dissipative mode. We also observed the formation of a dark state, a stationary soliton locked in place by the dissipation. This physics is captured by the effective description as well thanks to its preservation of the $U(1)$ symmetry.
Beyond the classical limit, in the truncated Wigner approximation, swichting between the metastable states is observed and quantified by the Liouvillian gap, the inverse of the asymptotic decay rate. From this we can conclude that the incoherent pumping model underestimates the fluctuations.

\begin{acknowledgements}
The authors would like to acknowledge fruitful discussions with H. Ott, A. Pelster, C. Mink, F. Minganti and L. Gravina. This work was supported by the FWO-Vlaanderen, project nr. 39532. Some of the computational resources and services used in this work were provided by the HPC core facility CalcUA of the Universiteit Antwerpen, and VSC (Flemish Supercomputer Center), funded by the Research Foundation - Flanders (FWO) and the Flemish Government.
\end{acknowledgements}

\bibliographystyle{apsrev4-2.bst}
\bibliography{Manuscript_I}

\begin{thebibliography}{56}%
\makeatletter
\providecommand \@ifxundefined [1]{%
 \@ifx{#1\undefined}
}%
\providecommand \@ifnum [1]{%
 \ifnum #1\expandafter \@firstoftwo
 \else \expandafter \@secondoftwo
 \fi
}%
\providecommand \@ifx [1]{%
 \ifx #1\expandafter \@firstoftwo
 \else \expandafter \@secondoftwo
 \fi
}%
\providecommand \natexlab [1]{#1}%
\providecommand \enquote  [1]{``#1''}%
\providecommand \bibnamefont  [1]{#1}%
\providecommand \bibfnamefont [1]{#1}%
\providecommand \citenamefont [1]{#1}%
\providecommand \href@noop [0]{\@secondoftwo}%
\providecommand \href [0]{\begingroup \@sanitize@url \@href}%
\providecommand \@href[1]{\@@startlink{#1}\@@href}%
\providecommand \@@href[1]{\endgroup#1\@@endlink}%
\providecommand \@sanitize@url [0]{\catcode `\\12\catcode `\$12\catcode
  `\&12\catcode `\#12\catcode `\^12\catcode `\_12\catcode `\%12\relax}%
\providecommand \@@startlink[1]{}%
\providecommand \@@endlink[0]{}%
\providecommand \url  [0]{\begingroup\@sanitize@url \@url }%
\providecommand \@url [1]{\endgroup\@href {#1}{\urlprefix }}%
\providecommand \urlprefix  [0]{URL }%
\providecommand \Eprint [0]{\href }%
\providecommand \doibase [0]{https://doi.org/}%
\providecommand \selectlanguage [0]{\@gobble}%
\providecommand \bibinfo  [0]{\@secondoftwo}%
\providecommand \bibfield  [0]{\@secondoftwo}%
\providecommand \translation [1]{[#1]}%
\providecommand \BibitemOpen [0]{}%
\providecommand \bibitemStop [0]{}%
\providecommand \bibitemNoStop [0]{.\EOS\space}%
\providecommand \EOS [0]{\spacefactor3000\relax}%
\providecommand \BibitemShut  [1]{\csname bibitem#1\endcsname}%
\let\auto@bib@innerbib\@empty
\bibitem [{\citenamefont {Diehl}\ \emph {et~al.}(2008)\citenamefont {Diehl},
  \citenamefont {Micheli}, \citenamefont {Kantian}, \citenamefont {Kraus},
  \citenamefont {Büchler},\ and\ \citenamefont {Zoller}}]{diehl_quantum_2008}%
  \BibitemOpen
  \bibfield  {author} {\bibinfo {author} {\bibfnamefont {S.}~\bibnamefont
  {Diehl}}, \bibinfo {author} {\bibfnamefont {A.}~\bibnamefont {Micheli}},
  \bibinfo {author} {\bibfnamefont {A.}~\bibnamefont {Kantian}}, \bibinfo
  {author} {\bibfnamefont {B.}~\bibnamefont {Kraus}}, \bibinfo {author}
  {\bibfnamefont {H.~P.}\ \bibnamefont {Büchler}},\ and\ \bibinfo {author}
  {\bibfnamefont {P.}~\bibnamefont {Zoller}},\ }\href
  {https://doi.org/10.1038/nphys1073} {\bibfield  {journal} {\bibinfo
  {journal} {Nature Phys}\ }\textbf {\bibinfo {volume} {4}},\ \bibinfo {pages}
  {878} (\bibinfo {year} {2008})}\BibitemShut {NoStop}%
\bibitem [{\citenamefont {Verstraete}\ \emph {et~al.}(2009)\citenamefont
  {Verstraete}, \citenamefont {Wolf},\ and\ \citenamefont
  {Ignacio~Cirac}}]{verstraete_quantum_2009}%
  \BibitemOpen
  \bibfield  {author} {\bibinfo {author} {\bibfnamefont {F.}~\bibnamefont
  {Verstraete}}, \bibinfo {author} {\bibfnamefont {M.~M.}\ \bibnamefont
  {Wolf}},\ and\ \bibinfo {author} {\bibfnamefont {J.}~\bibnamefont
  {Ignacio~Cirac}},\ }\href {https://doi.org/10.1038/nphys1342} {\bibfield
  {journal} {\bibinfo  {journal} {Nature Phys}\ }\textbf {\bibinfo {volume}
  {5}},\ \bibinfo {pages} {633} (\bibinfo {year} {2009})}\BibitemShut {NoStop}%
\bibitem [{\citenamefont {Barreiro}\ \emph {et~al.}(2011)\citenamefont
  {Barreiro}, \citenamefont {Müller}, \citenamefont {Schindler}, \citenamefont
  {Nigg}, \citenamefont {Monz}, \citenamefont {Chwalla}, \citenamefont
  {Hennrich}, \citenamefont {Roos}, \citenamefont {Zoller},\ and\ \citenamefont
  {Blatt}}]{barreiro_open-system_2011}%
  \BibitemOpen
  \bibfield  {author} {\bibinfo {author} {\bibfnamefont {J.~T.}\ \bibnamefont
  {Barreiro}}, \bibinfo {author} {\bibfnamefont {M.}~\bibnamefont {Müller}},
  \bibinfo {author} {\bibfnamefont {P.}~\bibnamefont {Schindler}}, \bibinfo
  {author} {\bibfnamefont {D.}~\bibnamefont {Nigg}}, \bibinfo {author}
  {\bibfnamefont {T.}~\bibnamefont {Monz}}, \bibinfo {author} {\bibfnamefont
  {M.}~\bibnamefont {Chwalla}}, \bibinfo {author} {\bibfnamefont
  {M.}~\bibnamefont {Hennrich}}, \bibinfo {author} {\bibfnamefont {C.~F.}\
  \bibnamefont {Roos}}, \bibinfo {author} {\bibfnamefont {P.}~\bibnamefont
  {Zoller}},\ and\ \bibinfo {author} {\bibfnamefont {R.}~\bibnamefont
  {Blatt}},\ }\href {https://doi.org/10.1038/nature09801} {\bibfield  {journal}
  {\bibinfo  {journal} {Nature}\ }\textbf {\bibinfo {volume} {470}},\ \bibinfo
  {pages} {486} (\bibinfo {year} {2011})}\BibitemShut {NoStop}%
\bibitem [{\citenamefont {Drummond}\ and\ \citenamefont
  {Walls}(1980)}]{drummond_quantum_1980}%
  \BibitemOpen
  \bibfield  {author} {\bibinfo {author} {\bibfnamefont {P.~D.}\ \bibnamefont
  {Drummond}}\ and\ \bibinfo {author} {\bibfnamefont {D.~F.}\ \bibnamefont
  {Walls}},\ }\href {https://doi.org/10.1088/0305-4470/13/2/034} {\bibfield
  {journal} {\bibinfo  {journal} {J. Phys. A: Math. Gen.}\ }\textbf {\bibinfo
  {volume} {13}},\ \bibinfo {pages} {725} (\bibinfo {year} {1980})}\BibitemShut
  {NoStop}%
\bibitem [{\citenamefont {Wouters}\ and\ \citenamefont
  {Carusotto}(2007)}]{wouters_excitations_2007}%
  \BibitemOpen
  \bibfield  {author} {\bibinfo {author} {\bibfnamefont {M.}~\bibnamefont
  {Wouters}}\ and\ \bibinfo {author} {\bibfnamefont {I.}~\bibnamefont
  {Carusotto}},\ }\href {https://doi.org/10.1103/PhysRevLett.99.140402}
  {\bibfield  {journal} {\bibinfo  {journal} {Phys. Rev. Lett.}\ }\textbf
  {\bibinfo {volume} {99}},\ \bibinfo {pages} {140402} (\bibinfo {year}
  {2007})}\BibitemShut {NoStop}%
\bibitem [{\citenamefont {Kessler}\ \emph {et~al.}(2012)\citenamefont
  {Kessler}, \citenamefont {Giedke}, \citenamefont {Imamoglu}, \citenamefont
  {Yelin}, \citenamefont {Lukin},\ and\ \citenamefont
  {Cirac}}]{kessler_dissipative_2012}%
  \BibitemOpen
  \bibfield  {author} {\bibinfo {author} {\bibfnamefont {E.~M.}\ \bibnamefont
  {Kessler}}, \bibinfo {author} {\bibfnamefont {G.}~\bibnamefont {Giedke}},
  \bibinfo {author} {\bibfnamefont {A.}~\bibnamefont {Imamoglu}}, \bibinfo
  {author} {\bibfnamefont {S.~F.}\ \bibnamefont {Yelin}}, \bibinfo {author}
  {\bibfnamefont {M.~D.}\ \bibnamefont {Lukin}},\ and\ \bibinfo {author}
  {\bibfnamefont {J.~I.}\ \bibnamefont {Cirac}},\ }\href
  {https://doi.org/10.1103/PhysRevA.86.012116} {\bibfield  {journal} {\bibinfo
  {journal} {Phys. Rev. A}\ }\textbf {\bibinfo {volume} {86}},\ \bibinfo
  {pages} {012116} (\bibinfo {year} {2012})}\BibitemShut {NoStop}%
\bibitem [{\citenamefont {Kordas}\ \emph {et~al.}(2012)\citenamefont {Kordas},
  \citenamefont {Wimberger},\ and\ \citenamefont
  {Witthaut}}]{kordas_dissipation-induced_2012}%
  \BibitemOpen
  \bibfield  {author} {\bibinfo {author} {\bibfnamefont {G.}~\bibnamefont
  {Kordas}}, \bibinfo {author} {\bibfnamefont {S.}~\bibnamefont {Wimberger}},\
  and\ \bibinfo {author} {\bibfnamefont {D.}~\bibnamefont {Witthaut}},\ }\href
  {https://doi.org/10.1209/0295-5075/100/30007} {\bibfield  {journal} {\bibinfo
   {journal} {EPL}\ }\textbf {\bibinfo {volume} {100}},\ \bibinfo {pages}
  {30007} (\bibinfo {year} {2012})}\BibitemShut {NoStop}%
\bibitem [{\citenamefont {Casteels}\ \emph {et~al.}(2016)\citenamefont
  {Casteels}, \citenamefont {Storme}, \citenamefont {Le~Boité},\ and\
  \citenamefont {Ciuti}}]{casteels_power_2016}%
  \BibitemOpen
  \bibfield  {author} {\bibinfo {author} {\bibfnamefont {W.}~\bibnamefont
  {Casteels}}, \bibinfo {author} {\bibfnamefont {F.}~\bibnamefont {Storme}},
  \bibinfo {author} {\bibfnamefont {A.}~\bibnamefont {Le~Boité}},\ and\
  \bibinfo {author} {\bibfnamefont {C.}~\bibnamefont {Ciuti}},\ }\href
  {https://doi.org/10.1103/PhysRevA.93.033824} {\bibfield  {journal} {\bibinfo
  {journal} {Phys. Rev. A}\ }\textbf {\bibinfo {volume} {93}},\ \bibinfo
  {pages} {033824} (\bibinfo {year} {2016})}\BibitemShut {NoStop}%
\bibitem [{\citenamefont {Yanay}\ and\ \citenamefont
  {Clerk}(2018)}]{yanay_reservoir_2018}%
  \BibitemOpen
  \bibfield  {author} {\bibinfo {author} {\bibfnamefont {Y.}~\bibnamefont
  {Yanay}}\ and\ \bibinfo {author} {\bibfnamefont {A.~A.}\ \bibnamefont
  {Clerk}},\ }\href {https://doi.org/10.1103/PhysRevA.98.043615} {\bibfield
  {journal} {\bibinfo  {journal} {Phys. Rev. A}\ }\textbf {\bibinfo {volume}
  {98}},\ \bibinfo {pages} {043615} (\bibinfo {year} {2018})}\BibitemShut
  {NoStop}%
\bibitem [{\citenamefont {Huybrechts}\ and\ \citenamefont
  {Wouters}(2020)}]{huybrechts_dynamical_2020}%
  \BibitemOpen
  \bibfield  {author} {\bibinfo {author} {\bibfnamefont {D.}~\bibnamefont
  {Huybrechts}}\ and\ \bibinfo {author} {\bibfnamefont {M.}~\bibnamefont
  {Wouters}},\ }\href {https://doi.org/10.1103/PhysRevA.102.053706} {\bibfield
  {journal} {\bibinfo  {journal} {Phys. Rev. A}\ }\textbf {\bibinfo {volume}
  {102}},\ \bibinfo {pages} {053706} (\bibinfo {year} {2020})}\BibitemShut
  {NoStop}%
\bibitem [{\citenamefont {Wang}\ \emph {et~al.}(2020)\citenamefont {Wang},
  \citenamefont {Navarrete-Benlloch},\ and\ \citenamefont
  {Cai}}]{wang_pattern_2020}%
  \BibitemOpen
  \bibfield  {author} {\bibinfo {author} {\bibfnamefont {Z.}~\bibnamefont
  {Wang}}, \bibinfo {author} {\bibfnamefont {C.}~\bibnamefont
  {Navarrete-Benlloch}},\ and\ \bibinfo {author} {\bibfnamefont
  {Z.}~\bibnamefont {Cai}},\ }\href
  {https://doi.org/10.1103/PhysRevLett.125.115301} {\bibfield  {journal}
  {\bibinfo  {journal} {Phys. Rev. Lett.}\ }\textbf {\bibinfo {volume} {125}},\
  \bibinfo {pages} {115301} (\bibinfo {year} {2020})}\BibitemShut {NoStop}%
\bibitem [{\citenamefont {Walls}\ and\ \citenamefont
  {Milburn}(2008)}]{walls_quantum_2008}%
  \BibitemOpen
  \bibfield  {author} {\bibinfo {author} {\bibfnamefont {D.~F.}\ \bibnamefont
  {Walls}}\ and\ \bibinfo {author} {\bibfnamefont {G.~J.}\ \bibnamefont
  {Milburn}},\ }\href@noop {} {\emph {\bibinfo {title} {Quantum optics}}},\
  \bibinfo {edition} {2nd}\ ed.\ (\bibinfo  {publisher} {Springer},\ \bibinfo
  {address} {Berlin},\ \bibinfo {year} {2008})\BibitemShut {NoStop}%
\bibitem [{\citenamefont {Carusotto}\ and\ \citenamefont
  {Ciuti}(2013)}]{carusotto_quantum_2013}%
  \BibitemOpen
  \bibfield  {author} {\bibinfo {author} {\bibfnamefont {I.}~\bibnamefont
  {Carusotto}}\ and\ \bibinfo {author} {\bibfnamefont {C.}~\bibnamefont
  {Ciuti}},\ }\href {https://doi.org/10.1103/RevModPhys.85.299} {\bibfield
  {journal} {\bibinfo  {journal} {Rev. Mod. Phys.}\ }\textbf {\bibinfo {volume}
  {85}},\ \bibinfo {pages} {299} (\bibinfo {year} {2013})}\BibitemShut
  {NoStop}%
\bibitem [{\citenamefont {Noh}\ and\ \citenamefont
  {Angelakis}(2017)}]{noh_quantum_2017}%
  \BibitemOpen
  \bibfield  {author} {\bibinfo {author} {\bibfnamefont {C.}~\bibnamefont
  {Noh}}\ and\ \bibinfo {author} {\bibfnamefont {D.~G.}\ \bibnamefont
  {Angelakis}},\ }\href {https://doi.org/10.1088/0034-4885/80/1/016401}
  {\bibfield  {journal} {\bibinfo  {journal} {Rep. Prog. Phys.}\ }\textbf
  {\bibinfo {volume} {80}},\ \bibinfo {pages} {016401} (\bibinfo {year}
  {2017})}\BibitemShut {NoStop}%
\bibitem [{\citenamefont {Verstraelen}\ and\ \citenamefont
  {Wouters}(2018)}]{verstraelen_gaussian_2018}%
  \BibitemOpen
  \bibfield  {author} {\bibinfo {author} {\bibfnamefont {W.}~\bibnamefont
  {Verstraelen}}\ and\ \bibinfo {author} {\bibfnamefont {M.}~\bibnamefont
  {Wouters}},\ }\href {https://doi.org/10.3390/app8091427} {\bibfield
  {journal} {\bibinfo  {journal} {Applied Sciences}\ }\textbf {\bibinfo
  {volume} {8}},\ \bibinfo {pages} {1427} (\bibinfo {year} {2018})}\BibitemShut
  {NoStop}%
\bibitem [{\citenamefont {Pieczarka}\ \emph {et~al.}(2020)\citenamefont
  {Pieczarka}, \citenamefont {Estrecho}, \citenamefont {Boozarjmehr},
  \citenamefont {Bleu}, \citenamefont {Steger}, \citenamefont {West},
  \citenamefont {Pfeiffer}, \citenamefont {Snoke}, \citenamefont {Levinsen},
  \citenamefont {Parish}, \citenamefont {Truscott},\ and\ \citenamefont
  {Ostrovskaya}}]{pieczarka_observation_2020}%
  \BibitemOpen
  \bibfield  {author} {\bibinfo {author} {\bibfnamefont {M.}~\bibnamefont
  {Pieczarka}}, \bibinfo {author} {\bibfnamefont {E.}~\bibnamefont {Estrecho}},
  \bibinfo {author} {\bibfnamefont {M.}~\bibnamefont {Boozarjmehr}}, \bibinfo
  {author} {\bibfnamefont {O.}~\bibnamefont {Bleu}}, \bibinfo {author}
  {\bibfnamefont {M.}~\bibnamefont {Steger}}, \bibinfo {author} {\bibfnamefont
  {K.}~\bibnamefont {West}}, \bibinfo {author} {\bibfnamefont {L.~N.}\
  \bibnamefont {Pfeiffer}}, \bibinfo {author} {\bibfnamefont {D.~W.}\
  \bibnamefont {Snoke}}, \bibinfo {author} {\bibfnamefont {J.}~\bibnamefont
  {Levinsen}}, \bibinfo {author} {\bibfnamefont {M.~M.}\ \bibnamefont
  {Parish}}, \bibinfo {author} {\bibfnamefont {A.~G.}\ \bibnamefont
  {Truscott}},\ and\ \bibinfo {author} {\bibfnamefont {E.~A.}\ \bibnamefont
  {Ostrovskaya}},\ }\href {https://doi.org/10.1038/s41467-019-14243-6}
  {\bibfield  {journal} {\bibinfo  {journal} {Nat Commun}\ }\textbf {\bibinfo
  {volume} {11}},\ \bibinfo {pages} {429} (\bibinfo {year} {2020})}\BibitemShut
  {NoStop}%
\bibitem [{\citenamefont {Labouvie}\ \emph {et~al.}(2015)\citenamefont
  {Labouvie}, \citenamefont {Santra}, \citenamefont {Heun}, \citenamefont
  {Wimberger},\ and\ \citenamefont {Ott}}]{labouvie_negative_2015}%
  \BibitemOpen
  \bibfield  {author} {\bibinfo {author} {\bibfnamefont {R.}~\bibnamefont
  {Labouvie}}, \bibinfo {author} {\bibfnamefont {B.}~\bibnamefont {Santra}},
  \bibinfo {author} {\bibfnamefont {S.}~\bibnamefont {Heun}}, \bibinfo {author}
  {\bibfnamefont {S.}~\bibnamefont {Wimberger}},\ and\ \bibinfo {author}
  {\bibfnamefont {H.}~\bibnamefont {Ott}},\ }\href
  {https://doi.org/10.1103/PhysRevLett.115.050601} {\bibfield  {journal}
  {\bibinfo  {journal} {Phys. Rev. Lett.}\ }\textbf {\bibinfo {volume} {115}},\
  \bibinfo {pages} {050601} (\bibinfo {year} {2015})}\BibitemShut {NoStop}%
\bibitem [{\citenamefont {Labouvie}\ \emph {et~al.}(2016)\citenamefont
  {Labouvie}, \citenamefont {Santra}, \citenamefont {Heun},\ and\ \citenamefont
  {Ott}}]{labouvie_bistability_2016}%
  \BibitemOpen
  \bibfield  {author} {\bibinfo {author} {\bibfnamefont {R.}~\bibnamefont
  {Labouvie}}, \bibinfo {author} {\bibfnamefont {B.}~\bibnamefont {Santra}},
  \bibinfo {author} {\bibfnamefont {S.}~\bibnamefont {Heun}},\ and\ \bibinfo
  {author} {\bibfnamefont {H.}~\bibnamefont {Ott}},\ }\href
  {https://doi.org/10.1103/PhysRevLett.116.235302} {\bibfield  {journal}
  {\bibinfo  {journal} {Phys. Rev. Lett.}\ }\textbf {\bibinfo {volume} {116}},\
  \bibinfo {pages} {235302} (\bibinfo {year} {2016})}\BibitemShut {NoStop}%
\bibitem [{\citenamefont {Benary}\ \emph {et~al.}(2022)\citenamefont {Benary},
  \citenamefont {Baals}, \citenamefont {Bernhart}, \citenamefont {Jiang},
  \citenamefont {Röhrle},\ and\ \citenamefont
  {Ott}}]{benary_experimental_2022}%
  \BibitemOpen
  \bibfield  {author} {\bibinfo {author} {\bibfnamefont {J.}~\bibnamefont
  {Benary}}, \bibinfo {author} {\bibfnamefont {C.}~\bibnamefont {Baals}},
  \bibinfo {author} {\bibfnamefont {E.}~\bibnamefont {Bernhart}}, \bibinfo
  {author} {\bibfnamefont {J.}~\bibnamefont {Jiang}}, \bibinfo {author}
  {\bibfnamefont {M.}~\bibnamefont {Röhrle}},\ and\ \bibinfo {author}
  {\bibfnamefont {H.}~\bibnamefont {Ott}},\ }\href
  {https://doi.org/10.1088/1367-2630/ac97b6} {\bibfield  {journal} {\bibinfo
  {journal} {New J. Phys.}\ }\textbf {\bibinfo {volume} {24}},\ \bibinfo
  {pages} {103034} (\bibinfo {year} {2022})}\BibitemShut {NoStop}%
\bibitem [{\citenamefont {Reeves}\ and\ \citenamefont
  {Davis}(2023)}]{reeves_bistability_2023}%
  \BibitemOpen
  \bibfield  {author} {\bibinfo {author} {\bibfnamefont {M.~T.}\ \bibnamefont
  {Reeves}}\ and\ \bibinfo {author} {\bibfnamefont {M.~J.}\ \bibnamefont
  {Davis}},\ }\href {https://doi.org/10.48550/arXiv.2102.02949} {\bibinfo
  {title} {Bistability and nonequilibrium condensation in a driven-dissipative
  {Josephson} array: a c-field model}} (\bibinfo {year} {2023}),\ \bibinfo
  {note} {arXiv:2102.02949}\BibitemShut {NoStop}%
\bibitem [{\citenamefont {Garbin}\ \emph {et~al.}(2022)\citenamefont {Garbin},
  \citenamefont {Giraldo}, \citenamefont {Peters}, \citenamefont {Broderick},
  \citenamefont {Spakman}, \citenamefont {Raineri}, \citenamefont {Levenson},
  \citenamefont {Rodriguez}, \citenamefont {Krauskopf},\ and\ \citenamefont
  {Yacomotti}}]{garbin_spontaneous_2022}%
  \BibitemOpen
  \bibfield  {author} {\bibinfo {author} {\bibfnamefont {B.}~\bibnamefont
  {Garbin}}, \bibinfo {author} {\bibfnamefont {A.}~\bibnamefont {Giraldo}},
  \bibinfo {author} {\bibfnamefont {K.~J.~H.}\ \bibnamefont {Peters}}, \bibinfo
  {author} {\bibfnamefont {N.~G.~R.}\ \bibnamefont {Broderick}}, \bibinfo
  {author} {\bibfnamefont {A.}~\bibnamefont {Spakman}}, \bibinfo {author}
  {\bibfnamefont {F.}~\bibnamefont {Raineri}}, \bibinfo {author} {\bibfnamefont
  {A.}~\bibnamefont {Levenson}}, \bibinfo {author} {\bibfnamefont {S.~R.~K.}\
  \bibnamefont {Rodriguez}}, \bibinfo {author} {\bibfnamefont {B.}~\bibnamefont
  {Krauskopf}},\ and\ \bibinfo {author} {\bibfnamefont {A.~M.}\ \bibnamefont
  {Yacomotti}},\ }\href {https://doi.org/10.1103/PhysRevLett.128.053901}
  {\bibfield  {journal} {\bibinfo  {journal} {Phys. Rev. Lett.}\ }\textbf
  {\bibinfo {volume} {128}},\ \bibinfo {pages} {053901} (\bibinfo {year}
  {2022})}\BibitemShut {NoStop}%
\bibitem [{\citenamefont {Giraldo}\ \emph {et~al.}(2020)\citenamefont
  {Giraldo}, \citenamefont {Krauskopf}, \citenamefont {Broderick},
  \citenamefont {Levenson},\ and\ \citenamefont
  {Yacomotti}}]{giraldo_driven-dissipative_2020}%
  \BibitemOpen
  \bibfield  {author} {\bibinfo {author} {\bibfnamefont {A.}~\bibnamefont
  {Giraldo}}, \bibinfo {author} {\bibfnamefont {B.}~\bibnamefont {Krauskopf}},
  \bibinfo {author} {\bibfnamefont {N.~G.~R.}\ \bibnamefont {Broderick}},
  \bibinfo {author} {\bibfnamefont {J.~A.}\ \bibnamefont {Levenson}},\ and\
  \bibinfo {author} {\bibfnamefont {A.~M.}\ \bibnamefont {Yacomotti}},\ }\href
  {https://doi.org/10.1088/1367-2630/ab7539} {\bibfield  {journal} {\bibinfo
  {journal} {New J. Phys.}\ }\textbf {\bibinfo {volume} {22}},\ \bibinfo
  {pages} {043009} (\bibinfo {year} {2020})}\BibitemShut {NoStop}%
\bibitem [{\citenamefont {Giraldo}\ \emph {et~al.}(2022)\citenamefont
  {Giraldo}, \citenamefont {Masson}, \citenamefont {Broderick},\ and\
  \citenamefont {Krauskopf}}]{giraldo_semiclassical_2022}%
  \BibitemOpen
  \bibfield  {author} {\bibinfo {author} {\bibfnamefont {A.}~\bibnamefont
  {Giraldo}}, \bibinfo {author} {\bibfnamefont {S.~J.}\ \bibnamefont {Masson}},
  \bibinfo {author} {\bibfnamefont {N.~G.~R.}\ \bibnamefont {Broderick}},\ and\
  \bibinfo {author} {\bibfnamefont {B.}~\bibnamefont {Krauskopf}},\ }\href
  {https://doi.org/10.1140/epjs/s11734-021-00416-2} {\bibfield  {journal}
  {\bibinfo  {journal} {Eur. Phys. J. Spec. Top.}\ }\textbf {\bibinfo {volume}
  {231}},\ \bibinfo {pages} {385} (\bibinfo {year} {2022})}\BibitemShut
  {NoStop}%
\bibitem [{\citenamefont {Witthaut}\ \emph {et~al.}(2011)\citenamefont
  {Witthaut}, \citenamefont {Trimborn}, \citenamefont {Hennig}, \citenamefont
  {Kordas}, \citenamefont {Geisel},\ and\ \citenamefont
  {Wimberger}}]{witthaut_beyond_2011}%
  \BibitemOpen
  \bibfield  {author} {\bibinfo {author} {\bibfnamefont {D.}~\bibnamefont
  {Witthaut}}, \bibinfo {author} {\bibfnamefont {F.}~\bibnamefont {Trimborn}},
  \bibinfo {author} {\bibfnamefont {H.}~\bibnamefont {Hennig}}, \bibinfo
  {author} {\bibfnamefont {G.}~\bibnamefont {Kordas}}, \bibinfo {author}
  {\bibfnamefont {T.}~\bibnamefont {Geisel}},\ and\ \bibinfo {author}
  {\bibfnamefont {S.}~\bibnamefont {Wimberger}},\ }\href
  {https://doi.org/10.1103/PhysRevA.83.063608} {\bibfield  {journal} {\bibinfo
  {journal} {Phys. Rev. A}\ }\textbf {\bibinfo {volume} {83}},\ \bibinfo
  {pages} {063608} (\bibinfo {year} {2011})}\BibitemShut {NoStop}%
\bibitem [{\citenamefont {Kordas}\ \emph {et~al.}(2015)\citenamefont {Kordas},
  \citenamefont {Witthaut},\ and\ \citenamefont
  {Wimberger}}]{kordas_non-equilibrium_2015}%
  \BibitemOpen
  \bibfield  {author} {\bibinfo {author} {\bibfnamefont {G.}~\bibnamefont
  {Kordas}}, \bibinfo {author} {\bibfnamefont {D.}~\bibnamefont {Witthaut}},\
  and\ \bibinfo {author} {\bibfnamefont {S.}~\bibnamefont {Wimberger}},\ }\href
  {https://doi.org/10.1002/andp.201400189} {\bibfield  {journal} {\bibinfo
  {journal} {Ann. Phys. (Berlin)}\ }\textbf {\bibinfo {volume} {527}},\
  \bibinfo {pages} {619} (\bibinfo {year} {2015})}\BibitemShut {NoStop}%
\bibitem [{\citenamefont {Barmettler}\ and\ \citenamefont
  {Kollath}(2011)}]{barmettler_controllable_2011}%
  \BibitemOpen
  \bibfield  {author} {\bibinfo {author} {\bibfnamefont {P.}~\bibnamefont
  {Barmettler}}\ and\ \bibinfo {author} {\bibfnamefont {C.}~\bibnamefont
  {Kollath}},\ }\href {https://doi.org/10.1103/PhysRevA.84.041606} {\bibfield
  {journal} {\bibinfo  {journal} {Phys. Rev. A}\ }\textbf {\bibinfo {volume}
  {84}},\ \bibinfo {pages} {041606} (\bibinfo {year} {2011})}\BibitemShut
  {NoStop}%
\bibitem [{\citenamefont {Kiefer-Emmanouilidis}\ and\ \citenamefont
  {Sirker}(2017)}]{kiefer-emmanouilidis_current_2017}%
  \BibitemOpen
  \bibfield  {author} {\bibinfo {author} {\bibfnamefont {M.}~\bibnamefont
  {Kiefer-Emmanouilidis}}\ and\ \bibinfo {author} {\bibfnamefont
  {J.}~\bibnamefont {Sirker}},\ }\href
  {https://doi.org/10.1103/PhysRevA.96.063625} {\bibfield  {journal} {\bibinfo
  {journal} {Phys. Rev. A}\ }\textbf {\bibinfo {volume} {96}},\ \bibinfo
  {pages} {063625} (\bibinfo {year} {2017})}\BibitemShut {NoStop}%
\bibitem [{\citenamefont {Kepesidis}\ and\ \citenamefont
  {Hartmann}(2012)}]{kepesidis_bose-hubbard_2012}%
  \BibitemOpen
  \bibfield  {author} {\bibinfo {author} {\bibfnamefont {K.~V.}\ \bibnamefont
  {Kepesidis}}\ and\ \bibinfo {author} {\bibfnamefont {M.~J.}\ \bibnamefont
  {Hartmann}},\ }\href {https://doi.org/10.1103/PhysRevA.85.063620} {\bibfield
  {journal} {\bibinfo  {journal} {Phys. Rev. A}\ }\textbf {\bibinfo {volume}
  {85}},\ \bibinfo {pages} {063620} (\bibinfo {year} {2012})}\BibitemShut
  {NoStop}%
\bibitem [{\citenamefont {Kordas}\ \emph {et~al.}(2013)\citenamefont {Kordas},
  \citenamefont {Wimberger},\ and\ \citenamefont
  {Witthaut}}]{kordas_decay_2013}%
  \BibitemOpen
  \bibfield  {author} {\bibinfo {author} {\bibfnamefont {G.}~\bibnamefont
  {Kordas}}, \bibinfo {author} {\bibfnamefont {S.}~\bibnamefont {Wimberger}},\
  and\ \bibinfo {author} {\bibfnamefont {D.}~\bibnamefont {Witthaut}},\ }\href
  {https://doi.org/10.1103/PhysRevA.87.043618} {\bibfield  {journal} {\bibinfo
  {journal} {Phys. Rev. A}\ }\textbf {\bibinfo {volume} {87}},\ \bibinfo
  {pages} {043618} (\bibinfo {year} {2013})}\BibitemShut {NoStop}%
\bibitem [{\citenamefont {Bychek}\ \emph {et~al.}(2019)\citenamefont {Bychek},
  \citenamefont {Muraev},\ and\ \citenamefont
  {Kolovsky}}]{bychek_probing_2019}%
  \BibitemOpen
  \bibfield  {author} {\bibinfo {author} {\bibfnamefont {A.~A.}\ \bibnamefont
  {Bychek}}, \bibinfo {author} {\bibfnamefont {P.~S.}\ \bibnamefont {Muraev}},\
  and\ \bibinfo {author} {\bibfnamefont {A.~R.}\ \bibnamefont {Kolovsky}},\
  }\href {https://doi.org/10.1103/PhysRevA.100.013610} {\bibfield  {journal}
  {\bibinfo  {journal} {Phys. Rev. A}\ }\textbf {\bibinfo {volume} {100}},\
  \bibinfo {pages} {013610} (\bibinfo {year} {2019})}\BibitemShut {NoStop}%
\bibitem [{\citenamefont {van Nieuwenburg}\ \emph {et~al.}(2018)\citenamefont
  {van Nieuwenburg}, \citenamefont {Malo}, \citenamefont {Daley},\ and\
  \citenamefont {Fischer}}]{van_nieuwenburg_dynamics_2018}%
  \BibitemOpen
  \bibfield  {author} {\bibinfo {author} {\bibfnamefont {E.}~\bibnamefont {van
  Nieuwenburg}}, \bibinfo {author} {\bibfnamefont {J.~Y.}\ \bibnamefont
  {Malo}}, \bibinfo {author} {\bibfnamefont {A.}~\bibnamefont {Daley}},\ and\
  \bibinfo {author} {\bibfnamefont {M.}~\bibnamefont {Fischer}},\ }\href
  {https://doi.org/10.1088/2058-9565/aa9a02} {\bibfield  {journal} {\bibinfo
  {journal} {Quantum Sci. Technol.}\ }\textbf {\bibinfo {volume} {3}},\
  \bibinfo {pages} {01LT02} (\bibinfo {year} {2018})}\BibitemShut {NoStop}%
\bibitem [{\citenamefont {Fröml}\ \emph {et~al.}(2020)\citenamefont {Fröml},
  \citenamefont {Muckel}, \citenamefont {Kollath}, \citenamefont
  {Chiocchetta},\ and\ \citenamefont {Diehl}}]{froml_ultracold_2020}%
  \BibitemOpen
  \bibfield  {author} {\bibinfo {author} {\bibfnamefont {H.}~\bibnamefont
  {Fröml}}, \bibinfo {author} {\bibfnamefont {C.}~\bibnamefont {Muckel}},
  \bibinfo {author} {\bibfnamefont {C.}~\bibnamefont {Kollath}}, \bibinfo
  {author} {\bibfnamefont {A.}~\bibnamefont {Chiocchetta}},\ and\ \bibinfo
  {author} {\bibfnamefont {S.}~\bibnamefont {Diehl}},\ }\href
  {https://doi.org/10.1103/PhysRevB.101.144301} {\bibfield  {journal} {\bibinfo
   {journal} {Phys. Rev. B}\ }\textbf {\bibinfo {volume} {101}},\ \bibinfo
  {pages} {144301} (\bibinfo {year} {2020})}\BibitemShut {NoStop}%
\bibitem [{\citenamefont {Brazhnyi}\ \emph {et~al.}(2009)\citenamefont
  {Brazhnyi}, \citenamefont {Konotop}, \citenamefont {Pérez-García},\ and\
  \citenamefont {Ott}}]{brazhnyi_dissipation-induced_2009}%
  \BibitemOpen
  \bibfield  {author} {\bibinfo {author} {\bibfnamefont {V.~A.}\ \bibnamefont
  {Brazhnyi}}, \bibinfo {author} {\bibfnamefont {V.~V.}\ \bibnamefont
  {Konotop}}, \bibinfo {author} {\bibfnamefont {V.~M.}\ \bibnamefont
  {Pérez-García}},\ and\ \bibinfo {author} {\bibfnamefont {H.}~\bibnamefont
  {Ott}},\ }\href {https://doi.org/10.1103/PhysRevLett.102.144101} {\bibfield
  {journal} {\bibinfo  {journal} {Phys. Rev. Lett.}\ }\textbf {\bibinfo
  {volume} {102}},\ \bibinfo {pages} {144101} (\bibinfo {year}
  {2009})}\BibitemShut {NoStop}%
\bibitem [{\citenamefont {Zezyulin}\ \emph {et~al.}(2012)\citenamefont
  {Zezyulin}, \citenamefont {Konotop}, \citenamefont {Barontini},\ and\
  \citenamefont {Ott}}]{zezyulin_macroscopic_2012}%
  \BibitemOpen
  \bibfield  {author} {\bibinfo {author} {\bibfnamefont {D.~A.}\ \bibnamefont
  {Zezyulin}}, \bibinfo {author} {\bibfnamefont {V.~V.}\ \bibnamefont
  {Konotop}}, \bibinfo {author} {\bibfnamefont {G.}~\bibnamefont {Barontini}},\
  and\ \bibinfo {author} {\bibfnamefont {H.}~\bibnamefont {Ott}},\ }\href
  {https://doi.org/10.1103/PhysRevLett.109.020405} {\bibfield  {journal}
  {\bibinfo  {journal} {Phys. Rev. Lett.}\ }\textbf {\bibinfo {volume} {109}},\
  \bibinfo {pages} {020405} (\bibinfo {year} {2012})}\BibitemShut {NoStop}%
\bibitem [{\citenamefont {Sels}\ and\ \citenamefont
  {Demler}(2020)}]{sels_thermal_2020}%
  \BibitemOpen
  \bibfield  {author} {\bibinfo {author} {\bibfnamefont {D.}~\bibnamefont
  {Sels}}\ and\ \bibinfo {author} {\bibfnamefont {E.}~\bibnamefont {Demler}},\
  }\href {https://doi.org/10.1016/j.aop.2019.168021} {\bibfield  {journal}
  {\bibinfo  {journal} {Annals of Physics}\ }\textbf {\bibinfo {volume}
  {412}},\ \bibinfo {pages} {168021} (\bibinfo {year} {2020})}\BibitemShut
  {NoStop}%
\bibitem [{\citenamefont {Bloch}\ \emph {et~al.}(2008)\citenamefont {Bloch},
  \citenamefont {Dalibard},\ and\ \citenamefont
  {Zwerger}}]{bloch_many-body_2008}%
  \BibitemOpen
  \bibfield  {author} {\bibinfo {author} {\bibfnamefont {I.}~\bibnamefont
  {Bloch}}, \bibinfo {author} {\bibfnamefont {J.}~\bibnamefont {Dalibard}},\
  and\ \bibinfo {author} {\bibfnamefont {W.}~\bibnamefont {Zwerger}},\ }\href
  {https://doi.org/10.1103/RevModPhys.80.885} {\bibfield  {journal} {\bibinfo
  {journal} {Rev. Mod. Phys.}\ }\textbf {\bibinfo {volume} {80}},\ \bibinfo
  {pages} {885} (\bibinfo {year} {2008})}\BibitemShut {NoStop}%
\bibitem [{\citenamefont {Breuer}\ and\ \citenamefont
  {Petruccione}(2002)}]{breuer_theory_2002}%
  \BibitemOpen
  \bibfield  {author} {\bibinfo {author} {\bibfnamefont {H.-P.}\ \bibnamefont
  {Breuer}}\ and\ \bibinfo {author} {\bibfnamefont {F.}~\bibnamefont
  {Petruccione}},\ }\href@noop {} {\emph {\bibinfo {title} {The theory of open
  quantum systems}}}\ (\bibinfo  {publisher} {Oxford University Press},\
  \bibinfo {address} {Oxford ; New York},\ \bibinfo {year} {2002})\BibitemShut
  {NoStop}%
\bibitem [{\citenamefont {Rackauckas}\ and\ \citenamefont
  {Nie}(2017)}]{rackauckas_differentialequationsjl_2017}%
  \BibitemOpen
  \bibfield  {author} {\bibinfo {author} {\bibfnamefont {C.}~\bibnamefont
  {Rackauckas}}\ and\ \bibinfo {author} {\bibfnamefont {Q.}~\bibnamefont
  {Nie}},\ }\href {https://doi.org/10.5334/jors.151} {\bibfield  {journal}
  {\bibinfo  {journal} {JORS}\ }\textbf {\bibinfo {volume} {5}},\ \bibinfo
  {pages} {15} (\bibinfo {year} {2017})}\BibitemShut {NoStop}%
\bibitem [{\citenamefont {Smerzi}\ \emph {et~al.}(1997)\citenamefont {Smerzi},
  \citenamefont {Fantoni}, \citenamefont {Giovanazzi},\ and\ \citenamefont
  {Shenoy}}]{smerzi_quantum_1997}%
  \BibitemOpen
  \bibfield  {author} {\bibinfo {author} {\bibfnamefont {A.}~\bibnamefont
  {Smerzi}}, \bibinfo {author} {\bibfnamefont {S.}~\bibnamefont {Fantoni}},
  \bibinfo {author} {\bibfnamefont {S.}~\bibnamefont {Giovanazzi}},\ and\
  \bibinfo {author} {\bibfnamefont {S.~R.}\ \bibnamefont {Shenoy}},\ }\href
  {https://doi.org/10.1103/PhysRevLett.79.4950} {\bibfield  {journal} {\bibinfo
   {journal} {Phys. Rev. Lett.}\ }\textbf {\bibinfo {volume} {79}},\ \bibinfo
  {pages} {4950} (\bibinfo {year} {1997})}\BibitemShut {NoStop}%
\bibitem [{\citenamefont {Meier}\ and\ \citenamefont
  {Zwerger}(2001)}]{meier_josephson_2001}%
  \BibitemOpen
  \bibfield  {author} {\bibinfo {author} {\bibfnamefont {F.}~\bibnamefont
  {Meier}}\ and\ \bibinfo {author} {\bibfnamefont {W.}~\bibnamefont
  {Zwerger}},\ }\href {https://doi.org/10.1103/PhysRevA.64.033610} {\bibfield
  {journal} {\bibinfo  {journal} {Phys. Rev. A}\ }\textbf {\bibinfo {volume}
  {64}},\ \bibinfo {pages} {033610} (\bibinfo {year} {2001})}\BibitemShut
  {NoStop}%
\bibitem [{\citenamefont {Albiez}\ \emph {et~al.}(2005)\citenamefont {Albiez},
  \citenamefont {Gati}, \citenamefont {Foelling}, \citenamefont {Hunsmann},
  \citenamefont {Cristiani},\ and\ \citenamefont
  {Oberthaler}}]{albiez_direct_2005}%
  \BibitemOpen
  \bibfield  {author} {\bibinfo {author} {\bibfnamefont {M.}~\bibnamefont
  {Albiez}}, \bibinfo {author} {\bibfnamefont {R.}~\bibnamefont {Gati}},
  \bibinfo {author} {\bibfnamefont {J.}~\bibnamefont {Foelling}}, \bibinfo
  {author} {\bibfnamefont {S.}~\bibnamefont {Hunsmann}}, \bibinfo {author}
  {\bibfnamefont {M.}~\bibnamefont {Cristiani}},\ and\ \bibinfo {author}
  {\bibfnamefont {M.~K.}\ \bibnamefont {Oberthaler}},\ }\href
  {https://doi.org/10.1103/PhysRevLett.95.010402} {\bibfield  {journal}
  {\bibinfo  {journal} {Phys. Rev. Lett.}\ }\textbf {\bibinfo {volume} {95}},\
  \bibinfo {pages} {010402} (\bibinfo {year} {2005})}\BibitemShut {NoStop}%
\bibitem [{\citenamefont {Seclì}\ \emph {et~al.}(2021)\citenamefont {Seclì},
  \citenamefont {Capone},\ and\ \citenamefont
  {Schirò}}]{secli_signatures_2021}%
  \BibitemOpen
  \bibfield  {author} {\bibinfo {author} {\bibfnamefont {M.}~\bibnamefont
  {Seclì}}, \bibinfo {author} {\bibfnamefont {M.}~\bibnamefont {Capone}},\
  and\ \bibinfo {author} {\bibfnamefont {M.}~\bibnamefont {Schirò}},\ }\href
  {https://doi.org/10.1088/1367-2630/ac04c8} {\bibfield  {journal} {\bibinfo
  {journal} {New J. Phys.}\ }\textbf {\bibinfo {volume} {23}},\ \bibinfo
  {pages} {063056} (\bibinfo {year} {2021})}\BibitemShut {NoStop}%
\bibitem [{\citenamefont {Pitaevskii}\ and\ \citenamefont
  {Stringari}(2016)}]{pitaevskii_bose-einstein_2016}%
  \BibitemOpen
  \bibfield  {author} {\bibinfo {author} {\bibfnamefont {L.}~\bibnamefont
  {Pitaevskii}}\ and\ \bibinfo {author} {\bibfnamefont {S.}~\bibnamefont
  {Stringari}},\ }\href@noop {} {\emph {\bibinfo {title} {Bose-{Einstein}
  {Condensation} and {Superfluidity}}}}\ (\bibinfo  {publisher} {Oxford
  University Press},\ \bibinfo {address} {Oxford, UK},\ \bibinfo {year}
  {2016})\BibitemShut {NoStop}%
\bibitem [{\citenamefont {Fischer}\ and\ \citenamefont
  {Wimberger}(2017)}]{fischer_models_2017}%
  \BibitemOpen
  \bibfield  {author} {\bibinfo {author} {\bibfnamefont {D.}~\bibnamefont
  {Fischer}}\ and\ \bibinfo {author} {\bibfnamefont {S.}~\bibnamefont
  {Wimberger}},\ }\href {https://doi.org/10.1002/andp.201600327} {\bibfield
  {journal} {\bibinfo  {journal} {Ann. Phys. (Berlin)}\ }\textbf {\bibinfo
  {volume} {529}},\ \bibinfo {pages} {1600327} (\bibinfo {year}
  {2017})}\BibitemShut {NoStop}%
\bibitem [{\citenamefont {Mink}\ \emph {et~al.}(2022)\citenamefont {Mink},
  \citenamefont {Pelster}, \citenamefont {Benary}, \citenamefont {Ott},\ and\
  \citenamefont {Fleischhauer}}]{mink_variational_2022}%
  \BibitemOpen
  \bibfield  {author} {\bibinfo {author} {\bibfnamefont {C.}~\bibnamefont
  {Mink}}, \bibinfo {author} {\bibfnamefont {A.}~\bibnamefont {Pelster}},
  \bibinfo {author} {\bibfnamefont {J.}~\bibnamefont {Benary}}, \bibinfo
  {author} {\bibfnamefont {H.}~\bibnamefont {Ott}},\ and\ \bibinfo {author}
  {\bibfnamefont {M.}~\bibnamefont {Fleischhauer}},\ }\href
  {https://doi.org/10.21468/SciPostPhys.12.2.051} {\bibfield  {journal}
  {\bibinfo  {journal} {SciPost Phys.}\ }\textbf {\bibinfo {volume} {12}},\
  \bibinfo {pages} {051} (\bibinfo {year} {2022})}\BibitemShut {NoStop}%
\bibitem [{\citenamefont {Kivshar}\ \emph {et~al.}(1994)\citenamefont
  {Kivshar}, \citenamefont {Królikowski},\ and\ \citenamefont
  {Chubykalo}}]{kivshar_dark_1994}%
  \BibitemOpen
  \bibfield  {author} {\bibinfo {author} {\bibfnamefont {Y.~S.}\ \bibnamefont
  {Kivshar}}, \bibinfo {author} {\bibfnamefont {W.}~\bibnamefont
  {Królikowski}},\ and\ \bibinfo {author} {\bibfnamefont {O.~A.}\ \bibnamefont
  {Chubykalo}},\ }\href {https://doi.org/10.1103/PhysRevE.50.5020} {\bibfield
  {journal} {\bibinfo  {journal} {Phys. Rev. E}\ }\textbf {\bibinfo {volume}
  {50}},\ \bibinfo {pages} {5020} (\bibinfo {year} {1994})}\BibitemShut
  {NoStop}%
\bibitem [{\citenamefont {Carr}\ \emph {et~al.}(2001)\citenamefont {Carr},
  \citenamefont {Brand}, \citenamefont {Burger},\ and\ \citenamefont
  {Sanpera}}]{carr_dark-soliton_2001}%
  \BibitemOpen
  \bibfield  {author} {\bibinfo {author} {\bibfnamefont {L.~D.}\ \bibnamefont
  {Carr}}, \bibinfo {author} {\bibfnamefont {J.}~\bibnamefont {Brand}},
  \bibinfo {author} {\bibfnamefont {S.}~\bibnamefont {Burger}},\ and\ \bibinfo
  {author} {\bibfnamefont {A.}~\bibnamefont {Sanpera}},\ }\href
  {https://doi.org/10.1103/PhysRevA.63.051601} {\bibfield  {journal} {\bibinfo
  {journal} {Phys. Rev. A}\ }\textbf {\bibinfo {volume} {63}},\ \bibinfo
  {pages} {051601} (\bibinfo {year} {2001})}\BibitemShut {NoStop}%
\bibitem [{\citenamefont {Mishmash}\ and\ \citenamefont
  {Carr}(2009)}]{mishmash_quantum_2009}%
  \BibitemOpen
  \bibfield  {author} {\bibinfo {author} {\bibfnamefont {R.~V.}\ \bibnamefont
  {Mishmash}}\ and\ \bibinfo {author} {\bibfnamefont {L.~D.}\ \bibnamefont
  {Carr}},\ }\href {https://doi.org/10.1103/PhysRevLett.103.140403} {\bibfield
  {journal} {\bibinfo  {journal} {Phys. Rev. Lett.}\ }\textbf {\bibinfo
  {volume} {103}},\ \bibinfo {pages} {140403} (\bibinfo {year}
  {2009})}\BibitemShut {NoStop}%
\bibitem [{\citenamefont {Mishmash}\ \emph {et~al.}(2009)\citenamefont
  {Mishmash}, \citenamefont {Danshita}, \citenamefont {Clark},\ and\
  \citenamefont {Carr}}]{mishmash_quantum_2009-1}%
  \BibitemOpen
  \bibfield  {author} {\bibinfo {author} {\bibfnamefont {R.~V.}\ \bibnamefont
  {Mishmash}}, \bibinfo {author} {\bibfnamefont {I.}~\bibnamefont {Danshita}},
  \bibinfo {author} {\bibfnamefont {C.~W.}\ \bibnamefont {Clark}},\ and\
  \bibinfo {author} {\bibfnamefont {L.~D.}\ \bibnamefont {Carr}},\ }\href
  {https://doi.org/10.1103/PhysRevA.80.053612} {\bibfield  {journal} {\bibinfo
  {journal} {Phys. Rev. A}\ }\textbf {\bibinfo {volume} {80}},\ \bibinfo
  {pages} {053612} (\bibinfo {year} {2009})}\BibitemShut {NoStop}%
\bibitem [{\citenamefont {Thorwart}\ and\ \citenamefont
  {Jung}(1997)}]{thorwart_dynamical_1997}%
  \BibitemOpen
  \bibfield  {author} {\bibinfo {author} {\bibfnamefont {M.}~\bibnamefont
  {Thorwart}}\ and\ \bibinfo {author} {\bibfnamefont {P.}~\bibnamefont
  {Jung}},\ }\href {https://doi.org/10.1103/PhysRevLett.78.2503} {\bibfield
  {journal} {\bibinfo  {journal} {Phys. Rev. Lett.}\ }\textbf {\bibinfo
  {volume} {78}},\ \bibinfo {pages} {2503} (\bibinfo {year}
  {1997})}\BibitemShut {NoStop}%
\bibitem [{\citenamefont {Thorwart}\ \emph {et~al.}(1998)\citenamefont
  {Thorwart}, \citenamefont {Reimann}, \citenamefont {Jung},\ and\
  \citenamefont {Fox}}]{thorwart_quantum_1998}%
  \BibitemOpen
  \bibfield  {author} {\bibinfo {author} {\bibfnamefont {M.}~\bibnamefont
  {Thorwart}}, \bibinfo {author} {\bibfnamefont {P.}~\bibnamefont {Reimann}},
  \bibinfo {author} {\bibfnamefont {P.}~\bibnamefont {Jung}},\ and\ \bibinfo
  {author} {\bibfnamefont {R.~F.}\ \bibnamefont {Fox}},\ }\href
  {https://doi.org/10.1016/S0301-0104(98)00128-1} {\bibfield  {journal}
  {\bibinfo  {journal} {Chemical Physics}\ }\textbf {\bibinfo {volume} {235}},\
  \bibinfo {pages} {61} (\bibinfo {year} {1998})}\BibitemShut {NoStop}%
\bibitem [{\citenamefont {Rodriguez}\ \emph {et~al.}(2017)\citenamefont
  {Rodriguez}, \citenamefont {Casteels}, \citenamefont {Storme}, \citenamefont
  {Carlon~Zambon}, \citenamefont {Sagnes}, \citenamefont {Le~Gratiet},
  \citenamefont {Galopin}, \citenamefont {Lemaître}, \citenamefont {Amo},
  \citenamefont {Ciuti},\ and\ \citenamefont {Bloch}}]{rodriguez_probing_2017}%
  \BibitemOpen
  \bibfield  {author} {\bibinfo {author} {\bibfnamefont {S.}~\bibnamefont
  {Rodriguez}}, \bibinfo {author} {\bibfnamefont {W.}~\bibnamefont {Casteels}},
  \bibinfo {author} {\bibfnamefont {F.}~\bibnamefont {Storme}}, \bibinfo
  {author} {\bibfnamefont {N.}~\bibnamefont {Carlon~Zambon}}, \bibinfo {author}
  {\bibfnamefont {I.}~\bibnamefont {Sagnes}}, \bibinfo {author} {\bibfnamefont
  {L.}~\bibnamefont {Le~Gratiet}}, \bibinfo {author} {\bibfnamefont
  {E.}~\bibnamefont {Galopin}}, \bibinfo {author} {\bibfnamefont
  {A.}~\bibnamefont {Lemaître}}, \bibinfo {author} {\bibfnamefont
  {A.}~\bibnamefont {Amo}}, \bibinfo {author} {\bibfnamefont {C.}~\bibnamefont
  {Ciuti}},\ and\ \bibinfo {author} {\bibfnamefont {J.}~\bibnamefont {Bloch}},\
  }\href {https://doi.org/10.1103/PhysRevLett.118.247402} {\bibfield  {journal}
  {\bibinfo  {journal} {Phys. Rev. Lett.}\ }\textbf {\bibinfo {volume} {118}},\
  \bibinfo {pages} {247402} (\bibinfo {year} {2017})}\BibitemShut {NoStop}%
\bibitem [{\citenamefont {Sinatra}\ \emph {et~al.}(2002)\citenamefont
  {Sinatra}, \citenamefont {Lobo},\ and\ \citenamefont
  {Castin}}]{sinatra_truncated_2002}%
  \BibitemOpen
  \bibfield  {author} {\bibinfo {author} {\bibfnamefont {A.}~\bibnamefont
  {Sinatra}}, \bibinfo {author} {\bibfnamefont {C.}~\bibnamefont {Lobo}},\ and\
  \bibinfo {author} {\bibfnamefont {Y.}~\bibnamefont {Castin}},\ }\href
  {https://doi.org/10.1088/0953-4075/35/17/301} {\bibfield  {journal} {\bibinfo
   {journal} {J. Phys. B: At. Mol. Opt. Phys.}\ }\textbf {\bibinfo {volume}
  {35}},\ \bibinfo {pages} {3599} (\bibinfo {year} {2002})}\BibitemShut
  {NoStop}%
\bibitem [{\citenamefont {Gardiner}\ and\ \citenamefont
  {Zoller}(2004)}]{gardiner_quantum_2004}%
  \BibitemOpen
  \bibfield  {author} {\bibinfo {author} {\bibfnamefont {C.~W.}\ \bibnamefont
  {Gardiner}}\ and\ \bibinfo {author} {\bibfnamefont {P.}~\bibnamefont
  {Zoller}},\ }\href@noop {} {\emph {\bibinfo {title} {Quantum noise}}},\
  \bibinfo {edition} {3rd}\ ed.\ (\bibinfo  {publisher} {Springer},\ \bibinfo
  {year} {2004})\BibitemShut {NoStop}%
\bibitem [{\citenamefont {Minganti}\ \emph {et~al.}(2018)\citenamefont
  {Minganti}, \citenamefont {Biella}, \citenamefont {Bartolo},\ and\
  \citenamefont {Ciuti}}]{minganti_spectral_2018}%
  \BibitemOpen
  \bibfield  {author} {\bibinfo {author} {\bibfnamefont {F.}~\bibnamefont
  {Minganti}}, \bibinfo {author} {\bibfnamefont {A.}~\bibnamefont {Biella}},
  \bibinfo {author} {\bibfnamefont {N.}~\bibnamefont {Bartolo}},\ and\ \bibinfo
  {author} {\bibfnamefont {C.}~\bibnamefont {Ciuti}},\ }\href
  {https://doi.org/10.1103/PhysRevA.98.042118} {\bibfield  {journal} {\bibinfo
  {journal} {Phys. Rev. A}\ }\textbf {\bibinfo {volume} {98}},\ \bibinfo
  {pages} {042118} (\bibinfo {year} {2018})}\BibitemShut {NoStop}%
\bibitem [{\citenamefont {Sinatra}\ \emph {et~al.}(2000)\citenamefont
  {Sinatra}, \citenamefont {Castin},\ and\ \citenamefont
  {Lobo}}]{sinatra_monte_2000}%
  \BibitemOpen
  \bibfield  {author} {\bibinfo {author} {\bibfnamefont {A.}~\bibnamefont
  {Sinatra}}, \bibinfo {author} {\bibfnamefont {Y.}~\bibnamefont {Castin}},\
  and\ \bibinfo {author} {\bibfnamefont {C.}~\bibnamefont {Lobo}},\ }\href
  {https://doi.org/10.1080/09500340008232186} {\bibfield  {journal} {\bibinfo
  {journal} {Journal of Modern Optics}\ }\textbf {\bibinfo {volume} {47}},\
  \bibinfo {pages} {2629} (\bibinfo {year} {2000})}\BibitemShut {NoStop}%
\end{thebibliography}%

\onecolumngrid
\appendix
\section{\label{AppSec:TWA} Truncated Wigner approximation}
A common approach to including quantum fluctuations for weakly interacting Bose gases in the quantum degenerate regime is the truncated Wigner approximation \cite{sinatra_truncated_2002,walls_quantum_2008}. This method, for a Bose-Hubbard chain of length $L$, is based on a description in $2L$-dimensional phase-space, where the Wigner distribution $W[A,A^*]$ is the analogue of the density matrix. An equation of motion for the Wigner function, equivalent to a master equation for the density matrix, can be derived thanks to a set of transformation rules \cite{gardiner_quantum_2004}:
\begin{equation}
\begin{split}
    \hat{a}_j\hat{\rho} &\leftrightarrow \left(\alpha_j + \frac{1}{2}\pdv{}{\alpha^*_j}\right)W[A,A^*],~~\hat{a}^\dagger_j\hat{\rho} \leftrightarrow \left(\alpha^*_j - \frac{1}{2}\pdv{}{\alpha_j}\right)W[A,A^*],\\
    \hat{\rho}\hat{a}_j &\leftrightarrow \left(\alpha_j - \frac{1}{2}\pdv{}{\alpha^*_j}\right)W[A,A^*],~~\hat{\rho}\hat{a}_j^\dagger \leftrightarrow \left(\alpha^*_j + \frac{1}{2}\pdv{}{\alpha_j}\right)W[A,A^*].
    \end{split}
\end{equation}
Applying these to the Lindblad master equation (Eq.~\eqref{Eq:LindbladMaster} in the main text) results in

\begin{equation}
    \begin{split}
&\pdv{W}{t} = -i\sum_j \Bigg(\pdv{}{\alpha_j}\bigg[J(\alpha_{j+1}+\alpha_{j-1}) - U\big[|\alpha_j|^2-1\big]\alpha_j+i\frac{\gamma_j}{2}\alpha_j\bigg] - c.c.\Bigg)W[A,A^*]\\ &+ \sum_j\Bigg(\frac{\gamma_j}{2}\pdv{}{\alpha_j^*}{\alpha_j} + \frac{iU}{4}\bigg[\frac{\partial^3}{\partial(\alpha_j^*)^2\partial\alpha_j}\alpha_j^*-\frac{\partial^3}{\partial\alpha_j^2\partial\alpha_j^*}\alpha_j\bigg]\Bigg)W[A,A^*].
\end{split}
\end{equation}

The approximation now consists of neglecting the third order derivatives. These terms scale with the inverse of the number of particles per site $1/n_j$, so for weak interactions $U$ and macroscopically large occupations the error of this approximation becomes relatively small. The remaining equation is a Focker-Planck equation, which can equivalently be written as a set of $L$ Langevin equations for the complex-valued phase-space variables:
\begin{equation}
    i\hbar\dv{\alpha_j}{t} = -J(\alpha_{j-1} + \alpha_{j+1}) + U\left(|\alpha_j|^2-1\right)\alpha_j - i\frac{\gamma_j}{2}\alpha_j + \sqrt{\gamma_j/2}\xi(t)
\end{equation}
The equations that are numerically solved in this approach are, apart from the stochastic term for $j=c$, identical to the GPE derived in mean-field. Yet, leading-order quantum fluctuations are taken into account by the non-deterministic nature of the initial conditions $\alpha_j(t=0)$. These values are sampled from the Wigner function representing the initial state of the system and subsequently time-evolved according to (\ref{Eq:BoseHubbardSDE}), yielding a so-called trajectory. Due to coupling to a reservoir an additional stochastic term comes in for the lossy site ($\gamma_j=\gamma\delta_{jc}$) with normalized complex Gaussian noise $\xi(t)$. Expectation values of Weyl-ordered products of operators correspond to moments of the Wigner function or, in terms of the time evolved stochastic variables $\alpha_j(t)$, to averages over large ensembles of numerically simulated trajectories. For the BHM this approach can actually be put on equal footing with Bogoliubov theory \cite{sinatra_monte_2000,mink_variational_2022}.

For the initial state we take our system at thermal equilibrium. To this end the total matter field at each site can be split up into components parallel and orthogonal to the condensate mode \cite{sinatra_truncated_2002}:
\begin{equation}
    \alpha_j(t=0) = a_{0}\phi^\parallel_j + \phi^\perp_j
\end{equation}
The condensate wave function $\phi^\parallel$ is a solution of the time-independent Gross-Pitaevskii equation of the closed system ($\gamma=0$) and is determined using imaginary time evolution. The non-condensed field is sampled considering the following expansion
\begin{equation}
    \phi^\perp_j = \sum_k b_k(x_j) u_ke^{ikx_j} + b^*_k(x_j) v^*_ke^{-ikx_j},
\end{equation}
were we sum over the eigenmodes $(u_k,v_k)$ of the discretized Bogoliubov operator with eigenenergies $\epsilon_k$. In the case of a BHM these are given by 
\begin{equation}
    u_k,v_k = \pm\left(\frac{\epsilon_k + Un}{2\omega_k}\pm \frac{1}{2}\right)^{1/2}
\end{equation}
with $\omega_k=\sqrt{\epsilon_k(\epsilon_k+2Un)}$, $\epsilon_k=4J\sin^2\left(ka/2\right)$ and $n=N/L$ the density of particles per site. The complex amplitudes for these eigenmodes are sampled from Gaussian distributions \cite{sinatra_truncated_2002}:
\begin{equation}
    P(b_k) = \frac{2}{\pi}\tanh\left(\frac{\epsilon_k}{2k_BT}\right)\exp\left[-2|b_k|^2\tanh\left(\frac{\epsilon_k}{2k_BT}\right)\right].
\end{equation}
We take the limit $T\rightarrow 0$ so that the amplitudes are sampled from a complex normal distribution. In order to conserve particle number at $t=0$ for each stochastic realization the number of non-condensed particles
\begin{equation}
    N_{exc} = \sum_k \left(\abs{u_k}^2 + \abs{v_k}^2\right)\left(b_k^*b_k-\frac{1}{2}\right)+\sum_k \abs{v_k}^2
\end{equation}
is subtracted from the fixed total number $N$ to find
\begin{equation}
    N_0=N-N_{exc}.
\end{equation}
For the start of each trajectory we then set $a_0=\sqrt{N_0+1/2}$, now also a stochastic variable.

\end{document}